\documentclass[aps,twocolumn,amsmath,amssymb,prd,superscriptaddress]{revtex4-2}
\usepackage{graphicx}
\usepackage{dcolumn}
\usepackage{mathrsfs}
\usepackage{pifont}
\usepackage{amsthm}
\usepackage{bm}
\usepackage{latexsym}
\usepackage[colorlinks=true,
linkcolor=blue,
urlcolor=blue,
citecolor=blue]{hyperref}
\usepackage[capitalise]{cleveref}
\usepackage{color}
\usepackage{booktabs} 

\newcommand{\ket}[1]{|#1\rangle}
\newcommand{\bra}[1]{\langle#1|}

\newtheorem{corollary}{Corollary}

\newtheorem{proposition}{Proposition}

\begin{document}
\title{Ascertaining higher-order quantum correlations in high energy physics} 
\author{Ao-Xiang Liu}
\affiliation{School of Physical Sciences, University of Chinese Academy of Sciences, 1 Yanqihu East Rd, Beijing 101408, China}
\author{Cong-Feng Qiao}
\email{qiaocf@ucas.ac.cn}
\affiliation{School of Physical Sciences, University of Chinese Academy of Sciences, 1 Yanqihu East Rd, Beijing 101408, China}
\affiliation{ICTP-AP, University of Chinese Academy of Sciences, Beijing 100190, China}
\date{\today}
\begin{abstract}
Nonlocality is a peculiar nature of quanta and it stands as an important quantum resource in application. Yet mere linear property of it, viz. the first order in moment, has been explored through various inequalities. Noticing the vast higher-order regime unexplored, in this study we investigate the higher-order quantum correlations in entangled hyperon-antihyperon  system, which may be generated massively in charmonium decays. A new type of Clauser-Horne inequality for statistical cumulants and central moments is formulated. We find that a significant violation of the third-order constraint, indicating the existence of higher-order correlation, exists in hyperon-antihyperon system and can be observed in high energy physics experiments, like BESIII and Belle II. Notably, the violation manifests more in higher energy systems of the $\Lambda\bar{\Lambda}$ pair against the kinematic contamination of timelike events.
\end{abstract}
\maketitle

\section{Introduction}

In their seminal 1935 paper \cite{einstein35}, Einstein, Podolsky, and Rosen (EPR) exposed the fundamental conflict between the assumptions of local realism and the completeness of the quantum mechanical description. This conceptual debate was transformed into a testable physical question by Bell, whose theorem provided a strict boundary between local hidden variable (LHV) theories and quantum mechanics \cite{bell64}. Since then, a rich hierarchy of EPR correlations, including entanglement, steering, and Bell nonlocality, has been established \cite{clauser69,freedman72,clauser74e,aspect82,wiseman07,cavalcanti09}, all of which are recognized as essential resources for quantum information processing \cite{OG09E,UR20Q,brunner14}. While experiments with photons have effectively ruled out LHV models \cite{freedman72,clauser74e,aspect82,branciard07,paterek07,groblacher07,branciard08,budroni22}, the verification of these fundamental principles using entangled massive particles at colliders remains a nascent frontier \cite{tornqvist81,li06,li09n,li10,shi20,fabbrichesi21,afik23}.

Verifying these principles in high-energy physics faces inherent challenges. High-energy decays are passive processes lacking active control over measurement settings \cite{bertlmann04}, and the absence of direct dichotomic observables complicates standard correlation functions \cite{afriat13}. Furthermore, Bell tests using Clauser-Horne-Shimony-Holt (CHSH) inequalities often suffer from the ``loophole'' induced by data renormalization \cite{hiesmayr15}. Recently, the generalized Clauser-Horne (CH) inequality \cite{qian20} addressed this renormalization issue by modeling hyperon weak decay within a generalized positive operator-valued measure (POVM) framework, which has been experimentally implemented in hyperon-antihyperon systems at BESIII \cite{ablikim25t}.

These tests remain limited by their reliance on first-order statistical moments (expectation values). The complete statistical information of a quantum measurement is, however, not fully captured by its mean. While the linear regime has been extensively explored through various inequalities, the higher-order regime remains virtually unexplored. 
Recent theoretical advances highlight that higher-order quantum correlations constitute a distinct quantum resource, potentially revealing nonlocal phenomena fundamentally different from standard Bell nonlocality \cite{yang23}. However, despite these insights, the experimental verification of such quantum correlations stays uncharted.

This deficiency is addressed in the present work by formulating a test of higher-order quantum correlations in entangled hyperon-antihyperon system. The focus is placed on the weak decays of entangled hyperon-antihyperon pairs produced in charmonium decays ($\chi_{c0}, \eta_c \to \Lambda\bar{\Lambda}$ and $J/\psi \to Y\bar{Y}$), where the hyperon weak decay functions as a self-analyzing polarimeter and is treated within the generalized POVM framework to account for decay parameters. 
Specifically, a hierarchy of criteria is derived, starting with the generalized CH inequality based on the second-order cumulant and extending to the third-order skewness constraint. The analysis reveals that the hyperon-antihyperon system exhibits a significant violation of this third-order constraint, serving as a signature of higher-order quantum correlations that is observable in high energy physics experiments like BESIII and Belle II. To ensure the validity of these tests in realistic collider environments, these algebraic bounds are rigorously modified to account for the unavoidable timelike separated background, a correction that confirms the specific robustness of nonlocality violations in the $\chi_{c0} \to \Lambda\bar{\Lambda}$ channel.
Furthermore, the analysis is extended to the fourth-order central moment. Its possible connection to state-independent contextuality is briefly discussed as a potential avenue for future investigation.

\section{Preliminaries}

Consider a quantum measurement described by a collection of measurement operators $\{M_m\}$ performed on a system in the state $\ket{\psi}$. The probability of obtaining the $m$-th outcome is given by $p_m=\bra{\psi} E_m\ket{\psi}$, where $E_m=M_m^\dagger M_m$ are positive operators satisfying the completeness relation $\sum_m E_m=I$. 

For a qubit (or spin-1/2) system, the measurement $M_m$ is typically implemented by an apparatus aligned along a direction $\boldsymbol{n}$, where each particle triggering the apparatus yields one of the dichotomic outcomes labeled $\pm$. The general POVM operators for such a system can be parametrized as:
\begin{align}
\label{eq:POVM}
E_{\pm}(\boldsymbol{a}) \equiv 
\frac{\xi^{(\pm)}\pm\alpha \boldsymbol{\sigma} \cdot \boldsymbol{a}}{2}\ .
\end{align}
Here, $\boldsymbol{\sigma}=\left(\sigma_x, \sigma_y, \sigma_z\right)$ is the vector of Pauli matrices. 
The parameters $\xi^{( \pm)}$ and $\alpha$ characterize the degrees of bias and unsharpness, respectively, defined by $\xi^{( \pm)}=1 \pm \eta$ with the constraint $|\eta \pm \alpha| \leq 1$. The general POVM in \cref{eq:POVM} reduces to the projective measurement when $\eta=0$, $|\alpha|=1$. Focusing on the ``$+$'' outcome, the probability of observing $+$ upon measuring along direction $\boldsymbol{a}$ is
\begin{align}
P_{+}(\boldsymbol{a}) \equiv\langle\psi| E_{+}(\boldsymbol{a})|\psi\rangle\ .
\end{align}
  
In an LHV theory, the measurement statistics are governed by a hidden parameter $\lambda$. In a bipartite system, the local observables $X$, $Y$ measured on subsystems $A$, $B$ are predetermined by hidden variable $\lambda$ as $X(\lambda)$, $Y(\lambda)$.
For a function $\mathcal{S}(X,Y)$ of local observables, its expectation value is given by:
\begin{align}
\label{eq:LHVavg}
\langle \mathcal{S}\rangle_{\text{LHV}}=\int  \mathcal{S}(X(\lambda),Y(\lambda))\rho (\lambda)\ \mathrm d\lambda\ ,
\end{align}
where $\rho(\lambda)$ is the normalized probability distribution of the hidden variable $\lambda$. 
According to \cref{eq:LHVavg}, the probability of obtaining a specific outcome in the LHV framework is:
\begin{align}
P_{\pm}(\boldsymbol{n})=\int P_{\pm}(\boldsymbol{n},\lambda)\rho (\lambda)\ \mathrm d\lambda\ ,
\end{align}
where $P_{\pm}(\boldsymbol{n},\lambda) \in [\frac{1\pm\eta-|\alpha|}{2},\frac{1\pm\eta+|\alpha|}{2}]$ represents the probability of the $\pm 1$ outcome conditioned on $\lambda$, bounded by the eigenvalues of the POVM operator. 
For a bipartite system composed of subsystems $A$ and $B$, the joint probability of outcomes $j$ and $k$ (with $j,k\in\{+,-\}$) along measurement directions $\boldsymbol{a}$ and $\boldsymbol{b}$ is:
\begin{align}
P_{jk}(\boldsymbol{a},\boldsymbol{b})=\int P_{j}(\boldsymbol{a},\lambda)P_{k}(\boldsymbol{b},\lambda)\rho (\lambda)\ \mathrm d\lambda\ .
\end{align}

Let us define the generalized Clauser-Horne (CH) functional for a bipartite system as follows:
\begin{align}
\label{eq:CH_functional}
\mathbf{B}_{\mathrm{CH}}^{jk}\equiv 
&P_{j k}(\boldsymbol{a}, \boldsymbol{b})-P_{j k}\left(\boldsymbol{a}, \boldsymbol{b}^{\prime}\right)+P_{j k}\left(\boldsymbol{a}^{\prime}, \boldsymbol{b}\right)+P_{j k}\left(\boldsymbol{a}^{\prime}, \boldsymbol{b}^{\prime}\right)\notag\\
& \quad-\left(1+k \eta_b\right) P_j\left(\boldsymbol{a}^{\prime}\right)-\left(1+j \eta_a\right) P_k(\boldsymbol{b})\notag \\
& \quad+\frac{\left(1+j \eta_a\right)\left(1+k \eta_b\right)-\left|\alpha_a\alpha_b\right|}{2}\ .
\end{align}
Here, $\eta_{a, b}$ and $\alpha_{a, b}$ are the bias and unsharpness parameters for subsystems $A$ and $B$, respectively. Without loss of generality, we set $j=k=+$ and omit the outcome superscripts in the subsequent analysis.

In quantum mechanics, the joint probability for a bipartite state $\rho$ is predicted by:
\begin{align}
P_{j k}(\boldsymbol{a}, \boldsymbol{b})=\operatorname{Tr}\left[\rho E_{j}(\boldsymbol{a}) \otimes E_{k}(\boldsymbol{b})\right]\ .
\end{align}
Consequently, the operator form of $\mathbf{B}_{\mathrm{CH}}$ in \cref{eq:CH} may be defined as the expectation value of the following operator: 
\begin{align}
\label{def:CHoperator}
\mathcal{B}_{\text{CH}} 
& \equiv E_+(\boldsymbol{a}) \otimes E_+(\boldsymbol{b})-E_+\left(\boldsymbol{a}\right) \otimes E_+\left(\boldsymbol{b}^{\prime}\right)\notag\\
& \quad +E_+\left(\boldsymbol{a}^{\prime}\right) \otimes E_+(\boldsymbol{b}) +E_+\left(\boldsymbol{a}^{\prime}\right) \otimes E_+\left(\boldsymbol{b}^{\prime}\right)\notag\\
& \quad -\left(1+ \eta_b\right) E_+\left(\boldsymbol{a}^{\prime}\right) \otimes I -\left(1+ \eta_a\right) I \otimes E_+(\boldsymbol{b})\notag\\
& \quad +\frac{\left(1+ \eta_a\right)\left(1+ \eta_b\right)-\left|\alpha_a\alpha_b\right|}{2} I\otimes I\ .
\end{align}
The expectation value of $\mathbf{B}_{\text{CH}}$ predicted by quantum theory is thus:
\begin{align}
\mathbf{B}_{\mathrm{CH}}=\langle \mathcal{B}_{\text{CH}}\rangle=\operatorname{Tr}\left(\rho \mathcal{B}_{\text{CH}}\right)\ .
\end{align}
For simplicity, the representative orthogonal measurement configuration ($\boldsymbol{a}\perp\boldsymbol{a}'$, $\boldsymbol{b}\perp\boldsymbol{b}'$) is adopted in this work, which is known to be optimal for revealing quantum nonlocality in spin-correlated systems \cite{cirel'son80,landau87}.


\section{Higher-order quantum correlations via cumulants and central moments} 
\label{sec:theory}
Standard Bell tests verify quantum nonlocality by constraining the first moment (mean value) of a Bell operator. However, the probability distribution of quantum measurement outcomes contains a complete hierarchy of statistical information that is often discarded in first-order tests. In this section, a theoretical framework that exploits this discarded information is established. 

\subsection{Statistical Moments and Cumulants}

Given a random variable $X$, the moment generating function takes the following form
\begin{align}
\label{eq:MGF}
\left\langle e^{s X}\right\rangle=\sum_{n=0}^{\infty}\left\langle X^n\right\rangle \frac{s^n}{n!}\ ,\quad s \in \mathbb{C}\ .
\end{align}
Here, the parameter $s$ is a complex number. Taking the natural logarithm of both sides of \cref{eq:MGF}, one obtains the cumulant generating function:
\begin{align}
K(s)=\ln \left\langle e^{s X}\right\rangle=\sum_{n=1}^{\infty} \kappa_n \frac{s^n}{n!}\ .
\end{align}
The coefficients $\kappa_n$ in the expansion are known as the cumulants of the random variable $X$. The first four order cumulants $\kappa_n$ are as follows:
\begin{align}
\kappa_1 (X) &= \langle X\rangle\ , (\text{Mean})\\
\kappa_2 (X) &= \mu_2(X)\ , (\text{Variance}) \\
\kappa_3 (X) &= \mu_3(X)\ , (\text{Skewness}) \label{eq:skewness}\\ 
\kappa_4 (X) &= \mu_4(X)-3[\mu_2(X)]^2\label{eq:kurtosis}\ . (\text{Kurtosis})
\end{align}
Here, the $n$-th order central moment $\mu_n(X)$ is defined as
\begin{align}
\mu_n(X)=\langle (X-\langle X\rangle)^n\rangle\ .
\end{align}

Cumulants are fundamental descriptors of the statistical properties of a random variable. While moments $\{\mu_n\}$ mix correlations of various orders, cumulants $\kappa_n$ isolate the pure, irreducible $n$-th order correlation structure, representing the part of the moment that cannot be explained by lower-order statistics. This property makes them indispensable in the study of correlations and non-Gaussianity.

\subsection{Higher-order quantum correlations in bipartite system}

Let $\mathcal{S}=\sum_{i,j}c_{ij}X_i\otimes Y_j$ be a joint measurement operator in a bipartite system with $c_{ij}\in\mathbb{C}$, then the statistical cumulant $\kappa_n(\mathcal{S})$ exists if the observables $X_i$, $Y_j$ have the moments up to order $n$ \cite{stuart10}. We present the following proposition:
\begin{proposition}
\label{prop:cumulant}
A bipartite system exhibits $n$-th order quantum correlations if it violates the following inequality:
\begin{align}
m\leq \kappa_n(\mathcal{S}) \leq M\ ,
\end{align}
where $m$ and $M$ are the lower and upper bounds of the $n$-th order cumulant $\kappa_n(\mathcal{S})$ imposed by classical theory.
\end{proposition}

As a demonstration of \cref{prop:cumulant}, we derive the generalized CH inequality from the nonnegativity of the second-order cumulant of the generalized CH operator $\mathcal{B}_{\mathrm{CH}}$ defined in \cref{def:CHoperator}.
In any LHV theory, the measurement outcomes are predetermined by a hidden variable $\lambda$ with distribution $\rho(\lambda)$. From \cref{eq:LHVavg}, the classical expectation value of $\mathcal{B}_{\text{CH}}$, is given by:
\begin{align}
\langle \mathcal{B}_{\mathrm{CH}}\rangle_{\mathrm{lhv}} = \int \mathcal{B}_{\text{CH}}(\lambda)\rho(\lambda) d\lambda\ ,
\end{align}
where $\mathcal{B}_{\text{CH}}(\lambda)$ represents the deterministic value of the operator for a given $\lambda$.

The critical distinction between quantum and classical descriptions lies in the algebraic properties of the operator. 
In quantum mechanics, the non-commutativity of local observables (e.g., $[\boldsymbol{\sigma}\cdot\boldsymbol{a}, \boldsymbol{\sigma}\cdot\boldsymbol{a}'] \neq 0$) allows the operator to span a larger spectrum. In contrast, within any LHV model, all local observables commute. Then, from the nonnegativity of the second-order cumulant of $\mathcal{B}_{\mathrm{CH}}$:
\begin{align}
\label{eq:k2CHnonneg}
\kappa_2(\mathcal{B}_{\mathrm{CH}}) \geq 0\ ,
\end{align}
we have the following fundamental corollary (see \cref{app:ProofCH} for details):
\begin{corollary}
\label{lemma:CH}
For any bipartite system composed of two subsystems $A$ and $B$, if the joint probabilities $p_{jk}(\boldsymbol{a},\boldsymbol{b})$ can be described by LHV theory, then the following inequalities hold:
\begin{align}
\label{eq:CH}
&-|\alpha_a\alpha_b|\leq \langle \mathcal{B}_{\mathrm{CH}} \rangle_{\mathrm{lhv}} \leq 0\ .
\end{align}
Here $\alpha_{a, b}$ are bias and unsharpness parameters on $A$ and $B$, defined in \cref{def:CHoperator}.
\end{corollary}
It is noteworthy that \cref{eq:CH} reproduces the standard CH inequality \cite{clauser74e} when $\alpha_a=\alpha_b=1$ (i.e., projective measurements without bias). While most previous works use only the upper bound of \cref{eq:CH} to test local realism, the lower bound is equally important in our analysis of higher-order quantum correlations.

As defined in \cref{eq:skewness}, the third-order cumulant (skewness) of $\mathcal{B}_{\mathrm{CH}}$ is
\begin{align}
\kappa_3(\mathcal{B}_{\mathrm{CH}})=\langle\mathcal{B}_{\mathrm{CH}}^3\rangle - 3\langle\mathcal{B}_{\mathrm{CH}}\rangle \langle\mathcal{B}_{\mathrm{CH}}^2\rangle + 2\langle\mathcal{B}_{\mathrm{CH}}\rangle^3\ .
\end{align}
The boundedness condition derived in \cref{lemma:CH} has profound implications for higher-order statistics. 
Exploiting the mathematical limits established for bounded random variables in \cite{egozcue12}, we have (see \cref{app:ProofCumulantBounds} for details):

\begin{corollary}
\label{theorem:CHcumulant}
For any bipartite system composed of two subsystems $A$ and $B$, the local realism imposes the following constraint on the third-order cumulant of the CH operator:
\begin{align}
\label{eq:k3_bound}
|\kappa_3(\mathcal{B}_{\text{CH}})| &\leq \frac{|\alpha_a\alpha_b|^3}{8}\ , 
\end{align}
Here, $\mathcal{B}_{\text{CH}}$ is defined in \cref{def:CHoperator}.
\end{corollary}

Violation of this skewness bound indicates the higher-order quantum correlation in the system, which provides a rigorous witness even when the \cref{eq:CH} holds.

\section{Formalism of Generalized POVM Measurement in Hyperon Decay}
\label{sec:formalism}

The weak decay of a hyperon $Y$ (e.g., $\Lambda$, $\Sigma^+$) exhibits a parity-violating angular distribution given by \cite{cronin63}:
\begin{align}
\frac{\mathrm{d} N_Y}{\mathrm{d} \Omega}=\frac{1}{4 \pi}\left(1+\alpha_Y \boldsymbol{P}_Y \cdot \boldsymbol{n}\right)\ ,
\end{align}
where $\boldsymbol{P}_Y$ is the hyperon polarization vector, $\boldsymbol{n}$ is the unit vector along the decay product's momentum, and $\alpha_Y$ is the weak decay parameter characterizing the asymmetry. In the helicity frame, where the $z$-axis aligns with the hyperon momentum and the $y$-axis is normal to the production plane, the probability of detecting the decay product along $\boldsymbol{n}$ is
\begin{align}
P_Y(\boldsymbol{n})=\frac{1}{2}\left(1+\alpha_Y |\boldsymbol{P}_Y| \cos\theta\right)\ ,
\end{align}
with $\theta$ denoting the angle between $\boldsymbol{P}_Y$ and $\boldsymbol{n}$.

To interpret this process through quantum measurement theory, we formalize the weak decay as a generalized POVM on the hyperon spin \cite{qian20}, extending the framework to mixed states. Without loss of generality, the decay $\Lambda \to p \pi^-$ serves as a representative example. With the initial $\Lambda$ spin state denoted by $\rho_\Lambda$ and the momentum space initialized in the ground state $\ket{\boldsymbol{g}_p}$, the total initial state is given by:
\begin{align}
\varrho_{\text{i}} = \rho_\Lambda \otimes |\boldsymbol{g}_p\rangle\langle \boldsymbol{g}_p|\ .
\end{align}
During the decay, the weak interaction operator $U$ couples the spin and momentum Hilbert spaces, evolving the system to
\begin{align} 
\varrho_{\text{f}} 
&=U (\rho_\Lambda \otimes |\boldsymbol{g}_p\rangle\langle \boldsymbol{g}_p|) U^\dagger\notag\\
&=\sum_{j,k \in \{+,-\}} (M_j(\boldsymbol{n}_p) \rho_\Lambda M_k^\dagger(\boldsymbol{n}_p)) \otimes |j\boldsymbol{n}_p\rangle\langle k\boldsymbol{n}_p| \ , 
\end{align}
where $\ket{\boldsymbol{n}_p}$ represents the proton momentum state. The Kraus operators $M_{\pm}(\boldsymbol{n}_p)$ are defined as
\begin{align}
M_\pm(\boldsymbol{n}_p)=\frac{1}{2(|S|^2+|P|^2)}\left(S \pm P \boldsymbol{\sigma}\cdot\boldsymbol{n}_p\right)\ ,
\end{align}
with $S$ and $P$ being the $s$- and $p$-wave decay amplitudes. The asymmetry parameter is identified as $\alpha_\Lambda=\frac{2 \operatorname{Re}(S^* P)}{|S|^2+|P|^2}$. 

Tracing out the spin degrees of freedom yields the reduced density matrix for the proton:
\begin{align}
\rho_{p}=\sum_{j,k \in \{+,-\}} \mathrm{Tr}[(M_j(\boldsymbol{n}_p) \rho_\Lambda M_k^\dagger(\boldsymbol{n}_p))] \otimes |j\boldsymbol{n}_p\rangle\langle k\boldsymbol{n}_p| \ . 
\end{align}
Consequently, a projective measurement of the proton momentum along $\boldsymbol{n}_p$ effectively implements a general POVM on the initial $\Lambda$ spin state, described by the elements
\begin{align}
\label{eq:POVMhyperon}
E_{\pm}(\boldsymbol{n}_p)
=\frac{1}{2}\left(I \pm \alpha_\Lambda \boldsymbol{\sigma}\cdot\boldsymbol{n}_p\right)\ ,
\end{align}
which satisfy the completeness relation $\sum_{\pm} E_{\pm}(\boldsymbol{n}_p)=I$. Here, $E_{\pm}(\boldsymbol{n}_p)=M_{\pm}^\dagger(\boldsymbol{n}_p) M_{\pm}(\boldsymbol{n}_p)$. Accordingly, the probability of capturing the proton in the direction $\boldsymbol{n}_p$ is predicted by quantum theory as 
\begin{align}
\label{eq:probability}
P_{\pm}(\boldsymbol{n}_p)
= \mathrm{Tr}[\rho_\Lambda E_\pm(\boldsymbol{n}_p)] = \frac{1}{2}(1 \pm \alpha_\Lambda \boldsymbol{P}_\Lambda \cdot \boldsymbol{n}_p)\ .
\end{align}

Extending this to the joint decay of an entangled hyperon-antihyperon pair (e.g., $\Lambda \to p \pi^-$ and $\bar{\Lambda} \to \bar{p} \pi^+$), the joint probability of detecting the decay products along directions $\boldsymbol{n}_p$ and $\boldsymbol{n}_{\bar{p}}$ is
\begin{align}
\label{eq:jointprob}
P_{jk}(\boldsymbol{n}_p,\boldsymbol{n}_{\bar{p}})
=\operatorname{Tr}\left[\rho_{\Lambda\bar{\Lambda}} E_{j}(\boldsymbol{n}_p) \otimes E_{k}(\boldsymbol{n}_{\bar{p}})\right]\ ,
\end{align}
where $j,k\in\{+,-\}$ and $\rho_{\Lambda\bar{\Lambda}}$ is the spin state of the entangled $\Lambda\bar{\Lambda}$ pair. This formalism is general and applies to other hyperon decays, such as $\Sigma^+ \to p \pi^0$, by substituting the appropriate decay parameter $\alpha_Y$.

Thus, hyperon weak decays provide a natural physical realization of general POVM measurements defined in \cref{eq:POVM} with bias $\eta=0$ and unsharpness $|\alpha_Y|<1$. Applying \cref{lemma:CH}, we obtain the CH inequality for this system:
\begin{align}
\label{eq:CHhyperon}
-|\alpha_Y\alpha_{\bar{Y}}|\leq\  
& \mathbf{B}_{\mathrm{CH}}\equiv P(\boldsymbol{n}_p, \boldsymbol{n}_{\bar{p}})-P\left(\boldsymbol{n}_p, \boldsymbol{n}_{\bar{p}}^{\prime}\right)+\notag\\
& P\left(\boldsymbol{n}_p^{\prime}, \boldsymbol{n}_{\bar{p}}\right)
 +P\left(\boldsymbol{n}_p^{\prime}, \boldsymbol{n}_{\bar{p}}^{\prime}\right)- P\left(\boldsymbol{n}_p^{\prime}\right)\notag\\
 & - P(\boldsymbol{n}_{\bar{p}})+\frac{1-|\alpha_Y||\alpha_{\bar{Y}}|}{2} \leq 0\ .
\end{align}
It is noteworthy that a rigorous test of local realism demands independence from assumptions intrinsic to quantum theory. Consequently, in the analysis of bipartite $Y\bar{Y}$ systems, the renormalization of spin correlation functions via the decay asymmetry parameter $\alpha_Y$ constitutes a logical circularity. In contrast, the criterion formulated in \cref{eq:CHhyperon} is constructed directly from measurement statistics, thereby circumventing this renormalization loophole.

In the subsequent analysis, we assume approximate CP conservation ($|\alpha_Y|=|\alpha_{\bar{Y}}|$). The relevant production and decay parameters for the $\chi_{c0}/\eta_c \to Y\bar{Y}$ and $J/\psi \to Y\bar{Y}$ processes are summarized in \cref{tab:combined_params}. 

\begin{table*}[htbp]
\centering
\footnotesize 
\caption{Summary of physical parameters for charmonium decays into ground-state octet hyperon pairs used in numerical simulations. The table lists: (1) ``Hyperon Decay" parameters including decay mode, asymmetry $\alpha_Y$, and branching fraction $\mathcal{B}_{dec}$; (2) ``Hyperon Velocity ($\beta$)" for parent particles $\eta_c, \chi_{c0}, J/\psi$, calculated from Particle Data Group masses \cite{navas24}; and (3) ``$J/\psi$ Production Dynamics" parameters, including branching fractions, polarization parameters, and corresponding references. $\mathcal{B}_{prod}$ is in units of $10^{-4}$ and $\mathcal{B}_{dec}$ in $\%$.}
\label{tab:combined_params}
\renewcommand{\arraystretch}{1.5} 
\setlength{\tabcolsep}{3pt} 
\begin{tabular}{llcc|ccc|cccc}
\toprule
\multicolumn{4}{c|}{Hyperon Decay Process} & \multicolumn{3}{c|}{Hyperon Velocity ($\boldsymbol{\beta}$)} & \multicolumn{4}{c}{$J/\psi$ Production Dynamics} \\
\cmidrule(lr){1-4} \cmidrule(lr){5-7} \cmidrule(lr){8-11}
Channel & Mode & $\alpha_Y$ & $\mathcal{B}_{dec}$ (\%) & $\eta_c$ & $\chi_{c0}$ &  $J/\psi$& $\mathcal{B}_{prod}$ ($10^{-4}$) & $\alpha_\psi$ & $\Delta\Phi$ (rad) & Refs \\
\midrule
$\Lambda\bar{\Lambda}$       & $\Lambda \to p\pi^-$     & $0.755(3)$  & $64$  & $0.664$ & $0.757$ & $0.693$ & $19.43(33)$ & $0.475(4)$   & $0.752(8)$   & \cite{ablikim19,ablikim22,ablikim17,ablikim22p} \\
$\Sigma^+\bar{\Sigma}^-$     & $\Sigma^+ \to p\pi^0$    & $-0.994(4)$ & $52$  & $0.604$ & $0.718$ & $0.640$ & $15.0(24)$  & $-0.508(7)$  & $-0.270(15)$ & \cite{ablikim08,ablikim20} \\
$\Xi^-\bar{\Xi}^+$           & $\Xi^- \to \Lambda\pi^-$ & $-0.379(4)$ & $100$ & $0.464$ & $0.633$ & $0.521$ & $9.7(8)$    & $0.586(16)$  & $1.213(49)$  & \cite{ablikim22,workman22} \\
$\Xi^0\bar{\Xi}^0$           & $\Xi^0 \to \Lambda\pi^0$ & $-0.375(3)$ & $96$  & $0.473$ & $0.638$ & $0.528$ & $11.65(4)$  & $0.514(16)$  & $1.168(26)$  & \cite{workman22,ablikim17s,ablikim23} \\
\bottomrule
\end{tabular}
\end{table*}

\section{Testing higher-order quantum correlations with entangled hyperon-antihyperon pairs}
In this section, the higher-order cumulant criteria are tested in entangled hyperon-antihyperon systems.

\subsection{Excluding the timelike background in experiments}
A genuine Bell test requires the entangled hyperon pairs to be spacelike separated, which cannot be guaranteed for each pair of hyperons produced in high-energy experiments. The timelike events may introduce local correlations that could mimic nonlocal quantum correlations, potentially invalidating the Bell test. Fortunately, one can obtain the fraction of space-like separation events over the total events. For a specific decay event, given the distances $x_1$ and $x_2$ traveled by the hyperon and anti-hyperon before decay, respectively, the condition for spacelike separation (in the rest frame of the mother particle) is:
\begin{align}
\frac{1}{k}\leq \frac{x_1}{x_2} \leq k\ .
\end{align}
Here $k=\frac{1+\beta_Y}{1-\beta_Y}$ and $\beta_Y=\frac{v_Y}{c}$ with $v_Y$ being the velocity of hyperon $Y$ and $c$ the speed of light. The fraction of spacelike separation events is given by \cite{tornqvist81} 
\begin{align}
f_s=\int_{0}^{\infty}e^{-x_2}\ \mathrm d x_2\int_{\frac{x_2}{k}}^{k x_2} e^{-x_1}\ \mathrm d x_1=\frac{k-1}{k+1}=\beta_Y\ .
\end{align}
While the joint probabilities of spacelike separated events must satisfy \cref{eq:CHhyperon} for local realism, i.e., upper bounded by zero, the timelike events may violate the inequality due to possible classical communication. Therefore, new bounds are needed.

\begin{figure}
\centering
\includegraphics[width=0.45\textwidth]{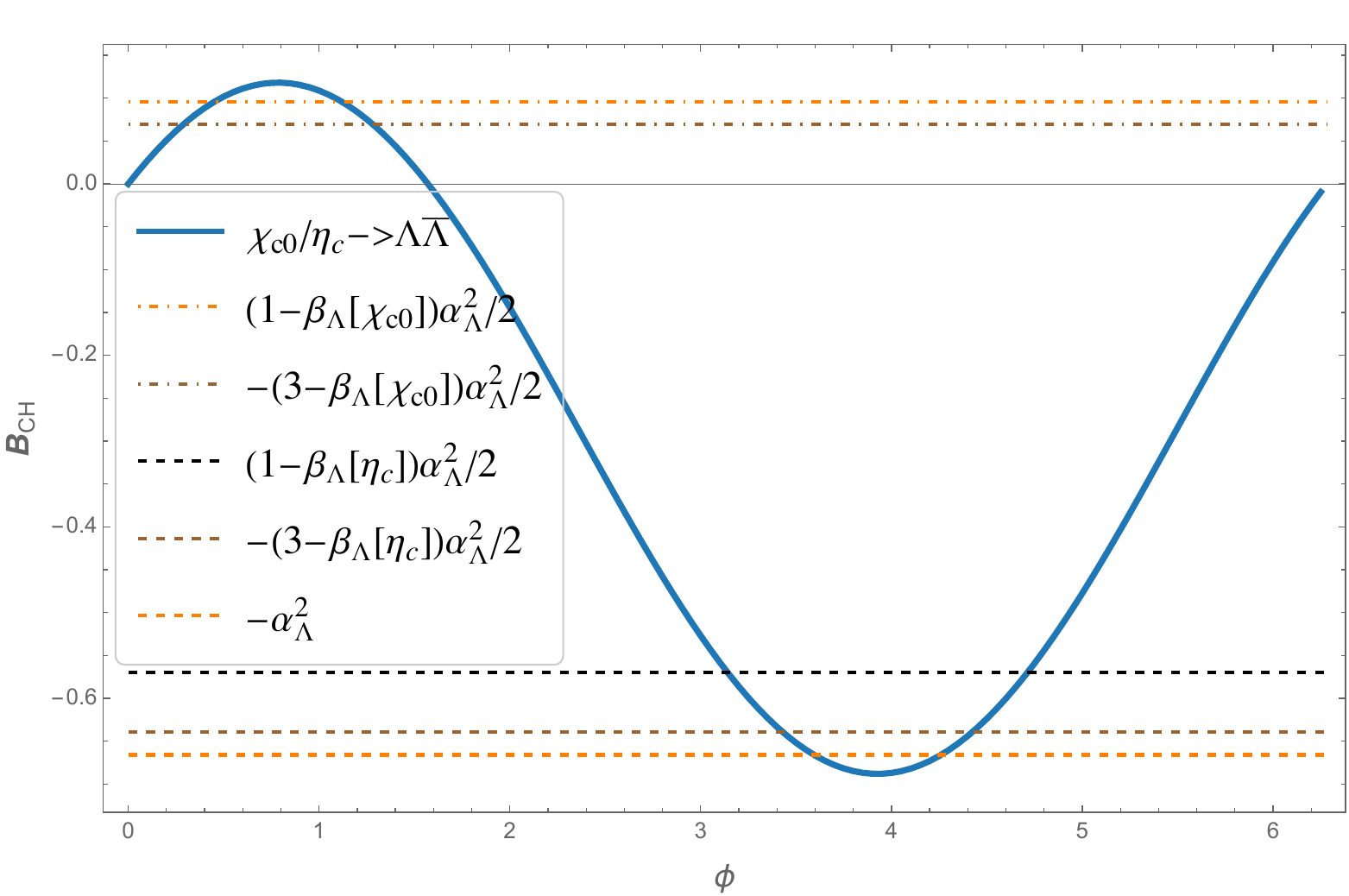}
\caption{Violation of the modified CH inequality \cref{eq:modifiedCH} for the process $\chi_{c0}/\eta_c \to \Lambda\bar{\Lambda}$. The solid blue curve represents the quantum mechanical prediction for the spin-singlet state.  The horizontal brown dot-dashed and dashed lines correspond to the upper and lower bounds in \cref{eq:modifiedCH} for $\chi_{c0}$, while the orange dot-dashed and dashed lines correspond to $\eta_c$. The black dashed line symbolizes the classical lower bound ($-\alpha_\Lambda^2$) in \cref{lemma:CH}.}
\label{fig:etacChic0LambdaBch}
\end{figure}

\begin{figure}
\centering
\includegraphics[width=0.45\textwidth]{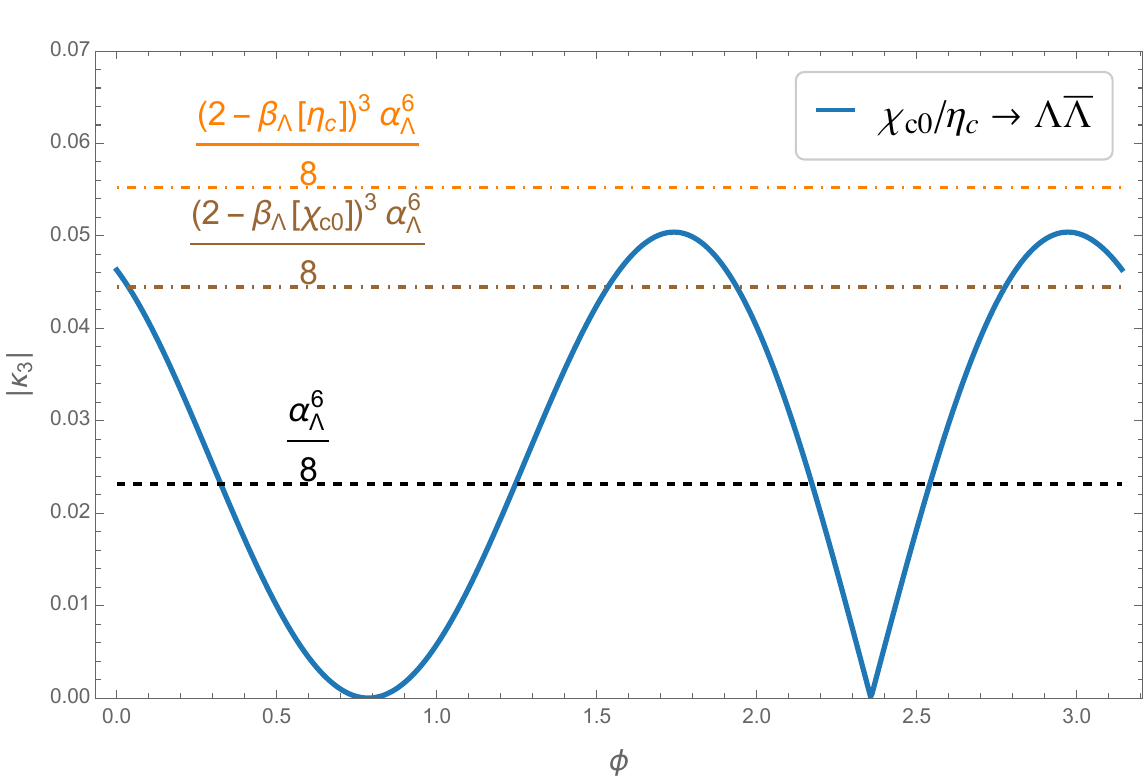}
\caption{Violation of the third-order cumulant criteria from \cref{theorem:CHcumulant} and \cref{corollary:hyperonCHcumulant} for the processes $\chi_{c0}/\eta_c \to \Lambda\bar{\Lambda}$. The solid blue curve represent the quantum prediction for the spin-singlet state in $\chi_{c0}/\eta_c$ decay. The horizontal orange and brown dot-dashed line symbolize the upper bounds in \cref{corollary:hyperonCHcumulant} for $\eta_c$ and $\chi_{c0}$ channels, respectively. The black dashed line indicates the classical upper bound defined in \cref{theorem:CHcumulant}.}
\label{fig:chic0LambdaSkew}
\end{figure}

As for the timelike events, the $\mathbf{B}_{\mathrm{CH}}$ in \cref{eq:CHhyperon} may reach its extreme values, that is,
\begin{align}
\label{eq:timelikeMax}
\max[\mathbf{B}_{\text{CH}}]= \frac{\alpha_Y^2}{2},\quad\text{or}\quad \min [\mathbf{B}_{\text{CH}}]= -\frac{3\alpha_Y^2}{2}
\end{align}
By considering the fraction of spacelike separation events $\beta_Y$, the modified upper and lower bounds of \cref{eq:CHhyperon} that accounts for the timelike background maybe respectively given by
\begin{align}
\label{eq:betaCHbound}
\beta_Y \cdot 0 + (1-\beta_Y) \cdot \frac{\alpha_Y^2}{2} = (1-\beta_Y) \frac{\alpha_Y^2}{2}\ ,
\end{align}
\begin{align}
\label{eq:betaCHlowerbound}
\beta_Y \cdot (-\alpha_Y^2) + (1-\beta_Y) \cdot \left(-\frac{3\alpha_Y^2}{2}\right) = -\left(3-\beta_Y\right) \frac{\alpha_Y^2}{2}\ .
\end{align}
Then, we have the modified CH inequality as \cite{qian20}
\begin{align}
\label{eq:modifiedCH}
-\left(3-\beta_Y\right) \frac{\alpha_Y^2}{2} \leq \mathbf{B}_{\text{CH}} \leq (1-\beta_Y) \frac{\alpha_Y^2}{2}\ .
\end{align}
This leads to a corresponding restriction for the third-order cumulant of $\mathcal{B}_{\text{CH}}$:
\begin{corollary}
\label{corollary:hyperonCHcumulant}
For a bipartite system, taking into account the fraction $\beta$ of spacelike separated events, local realism imposes the following constraint on the third-order cumulant of the CH operator:
\begin{align}
|\kappa_3(\mathcal{B}_{\text{CH}})| \leq 
\frac{(2-\beta)^3 \alpha^6}{8}\ .
\end{align}
\end{corollary}

\begin{figure}
\centering
\includegraphics[width=0.45\textwidth]{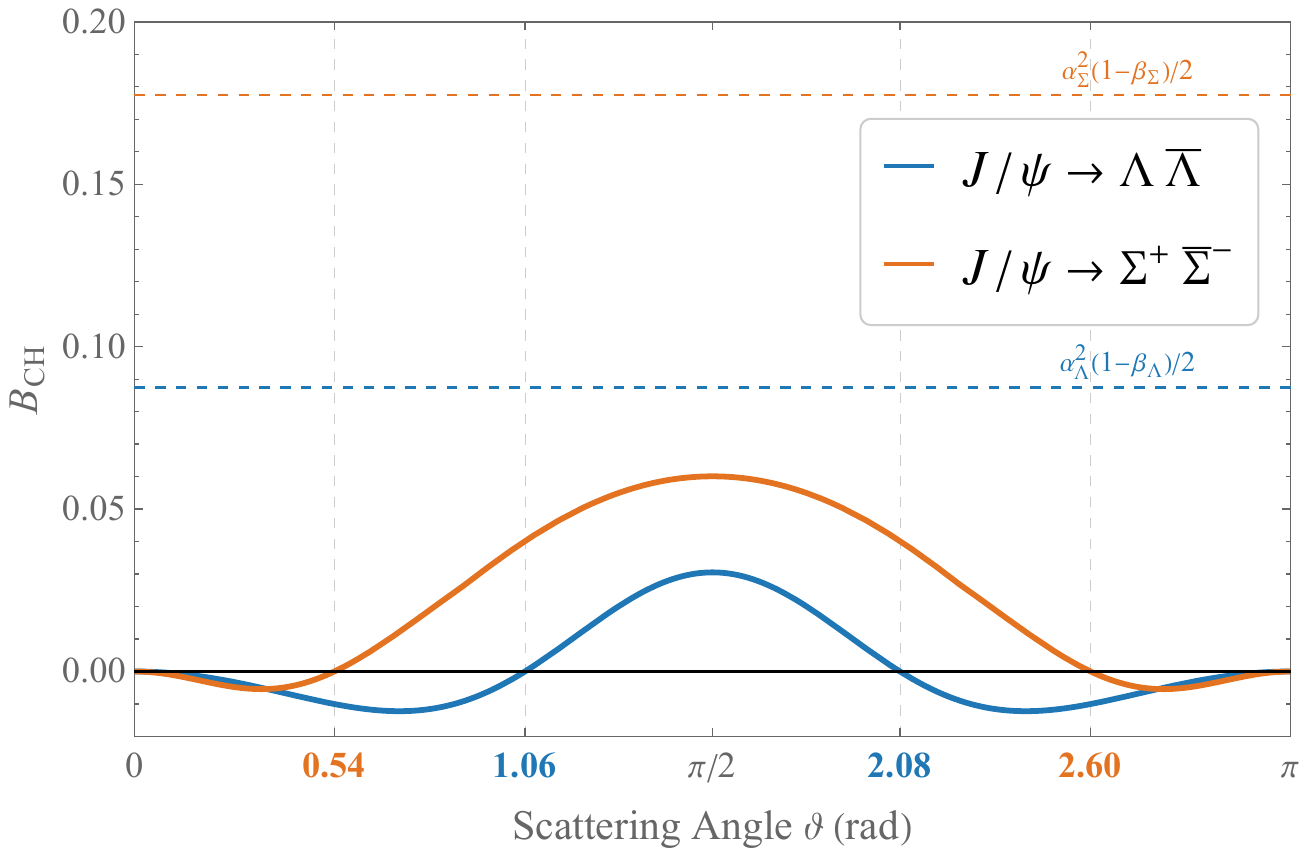}
\caption{Maximum value of $\langle\mathcal{B}_{\mathrm{CH}}\rangle$ versus scattering angle $\theta$ for $\Lambda\bar{\Lambda}$ (blue solid) and $\Sigma^+\bar{\Sigma}^-$ (brown solid) in $J/\psi$ decays. The classical upper bound ($0$) in \cref{eq:CHhyperon} is marked by the black line, while the modified upper bounds defined in \cref{eq:modifiedCH} are shown as dashed lines. Although the quantum predictions violate the former bound, they remain within the modified bounds across the entire range in terms of scattering angle.}
\label{fig:sigmaLambdaMeanNoviolation}
\end{figure}

\subsection{\texorpdfstring{$\chi_{c0}/\eta_c \to \Lambda\bar{\Lambda}$}{chi-c0/eta-c to Lambda Lambda-bar}}
For the process $\chi_{c0}/\eta_c \to \Lambda\bar{\Lambda}$, the spin state of the entangled $\Lambda\bar{\Lambda}$ pair is characterized by singlet state 
\begin{align}
|\psi_s\rangle=\frac{1}{\sqrt{2}}(|+ -\rangle - |- +\rangle)\ .
\end{align}
By adopting the measurement configuration as follows:
\begin{align}
\label{eq:measurementSetting1}
\boldsymbol{n}_p&=(1,0,0)\ ,\quad \boldsymbol{n}_p'=(   0,1,0)\ ,\notag\\
\boldsymbol{n}_{\bar{p}}&=(\cos\phi,\sin\phi,0)\ ,\quad \boldsymbol{n}^\prime_{\bar{p}}=(-\sin\phi,\cos\phi,0)\ ,
\end{align}
the quantum prediction for the CH quantity is
\begin{align}
\mathbf{B}_{\mathrm{CH}}=\langle\mathcal{B}_{\mathrm{CH}}\rangle=\frac{\alpha_\Lambda^2}{2}(\cos\phi+\sin\phi-1)\ .
\end{align}
The maximum value of $\mathbf{B}_{\text{CH}}$ is $\frac{\alpha_\Lambda^2}{2}(\sqrt{2}-1)$, which occurs at $\phi=\frac{\pi}{4}$. Thus, the violation of the modified CH inequality in \cref{eq:modifiedCH} requires $\beta_\Lambda > 2 - \sqrt{2} \approx 0.586$.
Since the spacelike fractions for $\chi_{c0}$ ($\beta_\Lambda[\chi_{c0}] \approx 0.757$) and $\eta_c$ ($\beta_\Lambda[\eta_c] \approx 0.664$) both exceed this threshold, the two processes can successfully demonstrate nonlocality via the generalized CH inequality see \cref{fig:etacChic0LambdaBch}. 

Using the same measurement settings in \cref{eq:measurementSetting1}, the skewness function is derived as:
\begin{align}
\kappa_3^{(\chi/\eta)} = -\frac{\sqrt{2}}{8} \alpha_\Lambda^6 \left[ \sin\left(\phi + \frac{\pi}{4}\right) - \sin\left(3\phi - \frac{\pi}{4}\right) \right]\ ,
\end{align}
which achieves its maximum value of $\frac{\sqrt{6}}{9} \alpha_\Lambda^6$ at $\arcsin\left(\frac{1}{\sqrt{3}}\right)-\frac{\pi}{4}$. The critical velocity threshold for the violation of the bound in \cref{corollary:hyperonCHcumulant} is $\beta_c = 2 - \sqrt[3]{8\sqrt{6}/9} \approx 0.704$. 

\begin{figure}
\centering
\includegraphics[width=0.45\textwidth]{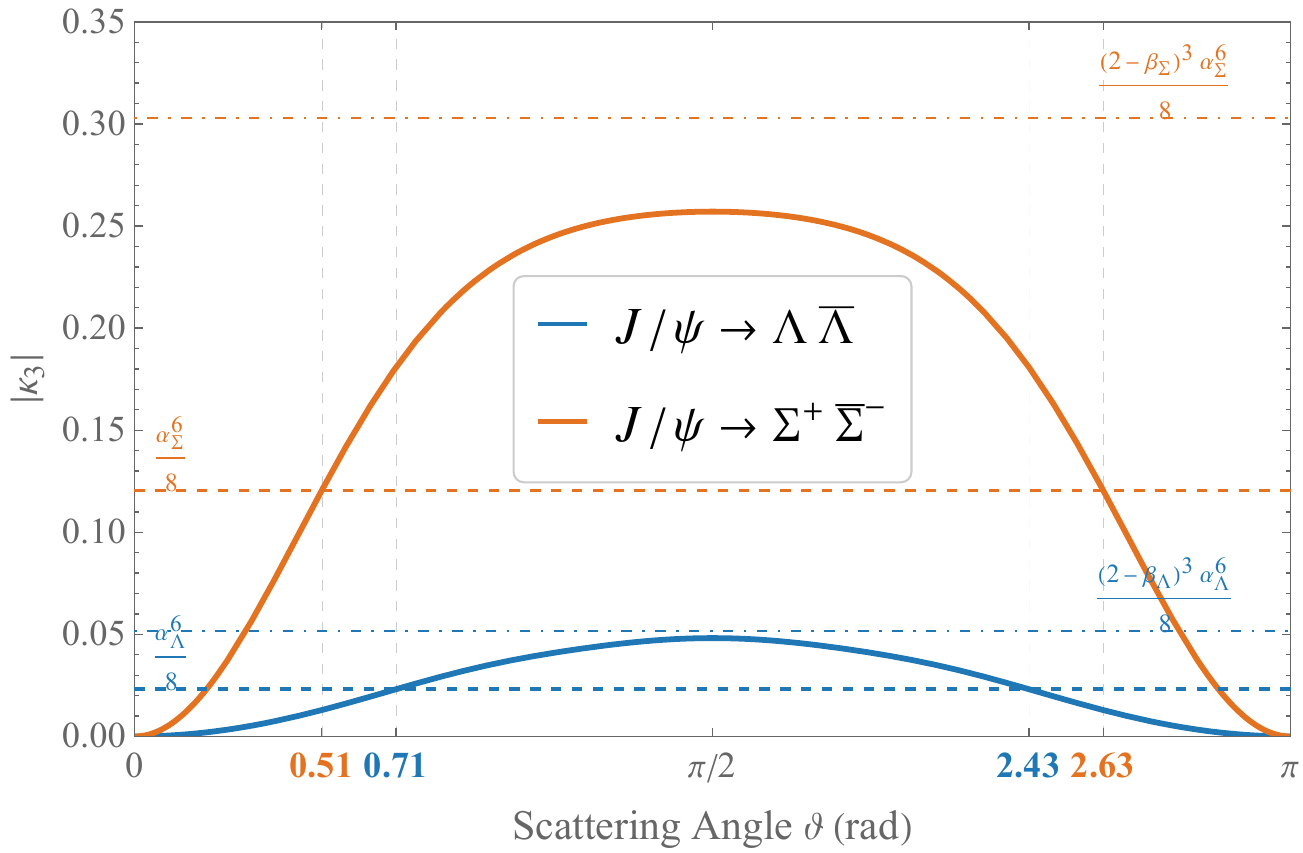}
\caption{Third-order cumulant $\kappa_3(\mathcal{B}_{\text{CH}})$ versus scattering angle $\vartheta$ for entangled $\Sigma^+\bar{\Sigma}^-$ and $\Lambda\bar{\Lambda}$ pairs produced in $J/\psi$ decays. The solid blue and brown curves represent the quantum predictions for $\Lambda\bar{\Lambda}$ and $\Sigma^+\bar{\Sigma}^-$, respectively. The horizontal dot-dashed lines indicate the modified upper bounds from \cref{corollary:hyperonCHcumulant}, while the dashed lines represent the upper bounds defined in \cref{theorem:CHcumulant} for $\Lambda\bar{\Lambda}$ (orange) and $\Sigma^+\bar{\Sigma}^-$ (brown).}
\label{fig:sigmaLambdaSkew}
\end{figure}

As illustrated in \cref{fig:chic0LambdaSkew}, the inequality in \cref{eq:k3_bound} is violated over a wide range of the parameter $\phi$. This indicates that the correlation terms in $\mathcal{B}_{\mathrm{CH}}$ induce a skewness significantly higher than classically admissible, reflecting the presence of higher-order dependencies. Although both the $\chi_{c0}\to \Lambda\bar{\Lambda}$ and $\eta_c\to \Lambda\bar{\Lambda}$ processes violate the bound derived in \cref{theorem:CHcumulant}, only the violation in the $\chi_{c0}$ channel persists under the modified bound given in \cref{corollary:hyperonCHcumulant}.

\subsection{\texorpdfstring{$J/\psi \to Y\bar{Y}$}{J/psi to Y Y-bar}}
In the BESIII experiment, large samples of entangled hyperon-antihyperon pairs are produced via $J/\psi$ decays. 
The spin state of the entangled $Y\bar{Y}$ pair is described by the general two-qubit density matrix:
\begin{align}
\label{eq:twoqubitstate}
\rho_{Y \bar{Y}}=& \frac{1}{4}[I \otimes I+\boldsymbol{P}_Y \cdot \boldsymbol{\sigma} \otimes I+I \otimes \boldsymbol{P}_{\bar{Y}} \cdot \boldsymbol{\sigma} \notag\\
&+\sum_{i,j} C_{i j} \sigma_i \otimes \sigma_j]\ .
\end{align}
Through local unitary operations \cite{yu07,wu24}, this state can be transformed into the standard symmetric two-qubit $X$-state:
\begin{align}
\label{eq:Xstate}
\rho_{Y \bar{Y}}
=\frac{1}{4}[I \otimes I+a\sigma_z \otimes I+I \otimes a\sigma_z+\sum_{i} t_i \sigma_i \otimes \sigma_i]\ .
\end{align}
The coefficients are determined by the decay kinematics:
\begin{align}
\label{eq:aP}
a=|\boldsymbol{P}_Y|=\frac{\gamma_\psi\sin\vartheta\cos\vartheta}{1+\alpha_\psi\cos^2\vartheta}\ ,
\end{align}
\begin{align}
\label{eq:t12}
t_{1,2}=\frac{1+\alpha_\psi\pm\sqrt{(1+\alpha_\psi\cos2\vartheta)^2-\gamma_\psi^2\sin^2 2\vartheta}}{2(1+\alpha_\psi\cos^2\vartheta)}\ ,
\end{align}
\begin{align}
t_3=\frac{-\alpha_\psi\sin^2\vartheta}{1+\alpha_\psi\cos^2\vartheta}\ ,
\end{align}
where $\gamma_\psi=\sqrt{1-\alpha_\psi^2} \sin \Delta \Phi$. Here, $\alpha_\psi \in [-1, 1]$ is the decay parameter of the vector charmonium $\psi$, $\Delta\Phi \in (-\pi, \pi]$ represents the relative phase of the form factors, and $\vartheta$ denotes the scattering angle of the hyperon $Y$ in the $J/\psi$ rest frame. The explicit values of $\alpha_\psi$ and $\Delta\Phi$ for various $J/\psi \to Y\bar{Y}$ channels are listed in \cref{tab:combined_params}.

\begin{figure}
\centering
\includegraphics[width=0.45\textwidth]{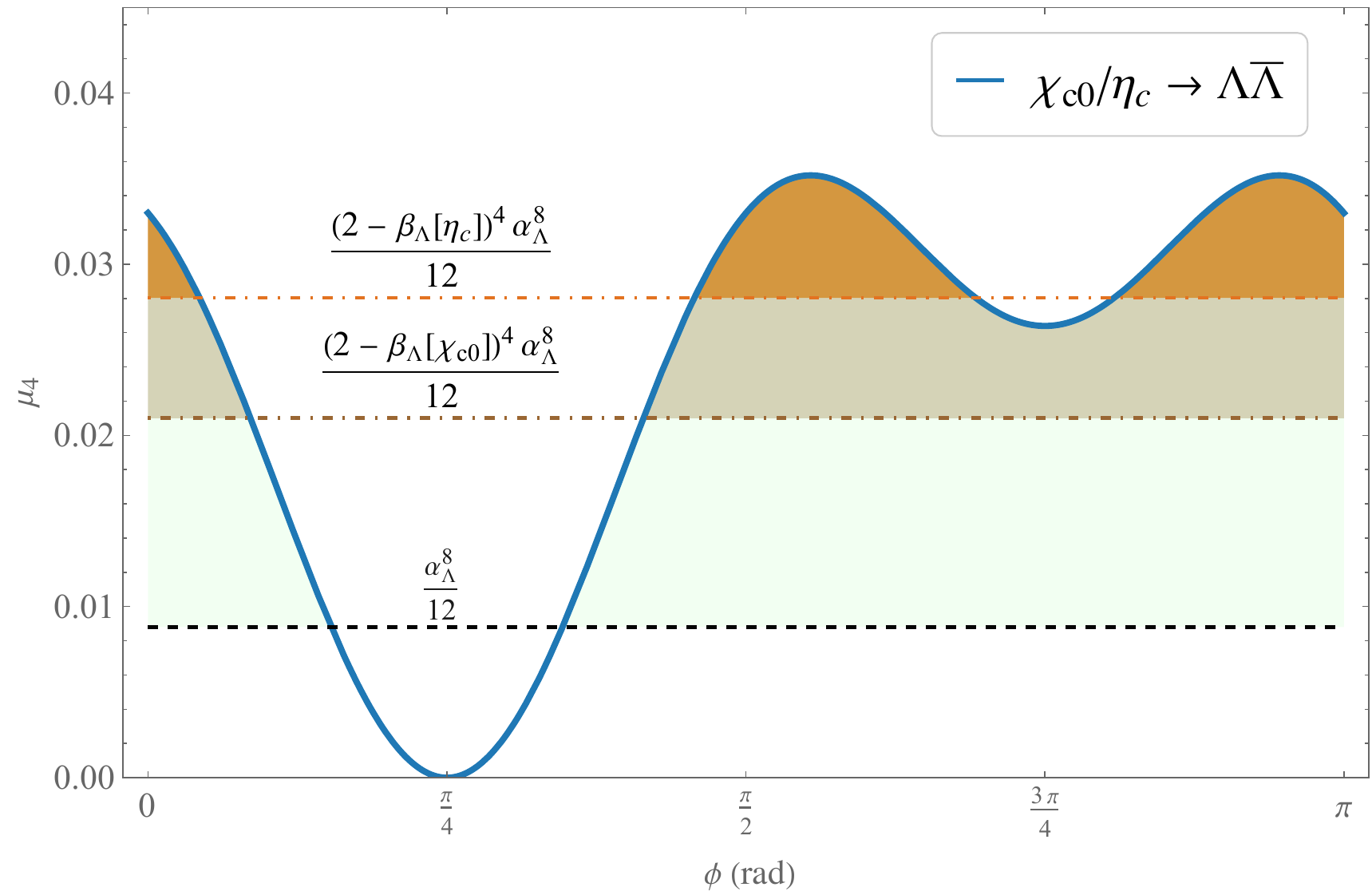}
\caption{Fourth-order central moment $\mu_4$ versus scanning angle $\phi$ for \texorpdfstring{$\chi_{c0}/\eta_c \rightarrow \Lambda\bar{\Lambda}$}{chi-c0/eta-c to Lambda Lambda-bar} decays. The quantum prediction is shown as a solid blue curve. The classical bound from \cref{eq:idealfourthCentrbound} is indicated by the black dashed line, while the timelike-modified bounds from \cref{eq:modifiedfourthCentrbound} for \texorpdfstring{$\chi_{c0}$}{chi-c0} and \texorpdfstring{$\eta_c$}{eta-c} are depicted by dot-dashed lines. Shaded regions highlight the violations of these classical limits.}
\label{fig:etacChic0FourthCentral}
\end{figure}

For entangled hyperon-antihyperon pairs $Y\bar{Y}$ produced in $J/\psi$ decays, the maximal expectation value of the CHSH operator is given by \cite{horodecki95}:
\begin{align}
2\sqrt{\lambda_1+\lambda_2}\ ,
\end{align}
where $\lambda_1$ and $\lambda_2$ ($\lambda_i \le 1$) denote the two largest eigenvalues of the correlation matrix $T=\mathrm{diag}(t_1,t_2,t_3)$. 
Correspondingly, the expectation value of the generalized CH operator $\mathcal{B}_{\text{CH}}$ (defined in \cref{def:CHoperator}) is bounded by the quantum mechanical limits:
\begin{align}
\label{eq:CHmax}
-\frac{\alpha_Y^2\sqrt{\lambda_1+\lambda_2}}{2}-\frac{1}{2}\alpha_Y^2\leq
\langle\mathcal{B}_{\text{CH}}\rangle
\leq\frac{\alpha_Y^2\sqrt{\lambda_1+\lambda_2}}{2}-\frac{1}{2}\alpha_Y^2\ .
\end{align}
Violations of the bounds in \cref{eq:CHmax} occur when:
\begin{align}
\label{eq:YYviolationcondition}
\frac{\alpha_Y^2\sqrt{\lambda_1+\lambda_2}}{2}-\frac{1}{2}\alpha_Y^2 & > (1-\beta_Y) \frac{\alpha_Y^2}{2}\ ,\\
-\frac{\alpha_Y^2\sqrt{\lambda_1+\lambda_2}}{2}-\frac{1}{2}\alpha_Y^2 & < -\left(3-\beta_Y\right) \frac{\alpha_Y^2}{2}\ .
\end{align}

\begin{figure}
\centering
\includegraphics[width=0.45\textwidth]{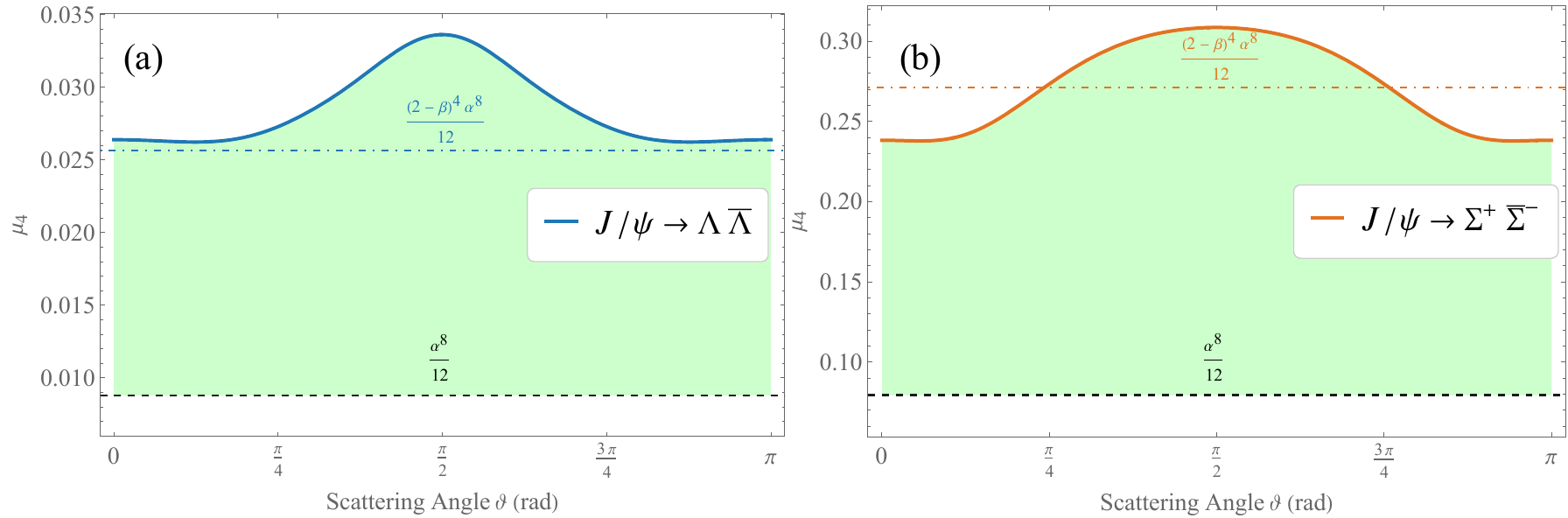}
\caption{The fourth-order central moment $\mu_4$ as a function of the scattering angle $\vartheta$ for entangled hyperon pairs produced in \texorpdfstring{$J/\psi$}{J/psi} decay. 
(a) The results for the \texorpdfstring{$\Sigma^+\bar{\Sigma}^-$}{Sigma+ Sigma-bar} pair. The solid blue curve represents the quantum prediction, while the blue dot-dashed line denotes the modified upper bound derived in \cref{eq:modifiedfourthCentrbound}. 
(b) The results for the \texorpdfstring{$\Lambda\bar{\Lambda}$}{Lambda Lambda-bar} pair. The solid orange curve represents the quantum prediction, and the orange dot-dashed line denotes the modified upper bound. 
In both panels, the black dashed lines indicate the ideal classical upper bound from \cref{eq:idealfourthCentrbound}.}
\label{fig:JpsiLsFourthCentral}
\end{figure}

The quantum predicted value $\langle\mathcal{B}_{\mathrm{CH}}\rangle$ depends on the scattering angle $\vartheta$, attaining its maximal values at $\vartheta=\pi/2$:
\begin{align}
\label{eq:CHmax_at_pi2}
\frac{\alpha_Y^2\sqrt{1+\alpha_\psi^2}}{2}-\frac{1}{2}\alpha_Y^2\ .
\end{align}
Therefore, violating the modified upper bound in \cref{eq:betaCHbound} requires the following kinematic condition:
\begin{align}
\beta_Y + \sqrt{1+\alpha_\psi^2} - 2 > 0\ .
\end{align}
In the relativistic limit $\beta_Y\to 1$, the CH inequalities in \cref{eq:CHhyperon} would indeed be violated by the hyperon pairs from $J/\psi$ decay. However, for the relevant hyperon channels ($Y = \Lambda, \Sigma^+, \Xi^0, \Xi^-$), this condition remains unsatisfied due to the moderate hyperon velocities $\beta_Y$ and the small magnitude of the decay parameter $\alpha_\psi$.
This is illustrated in \cref{fig:sigmaLambdaMeanNoviolation}, where the left-hand side of \cref{eq:YYviolationcondition} is plotted against $\vartheta$ for entangled $\Sigma^+\bar{\Sigma}^-$ and $\Lambda\bar{\Lambda}$ pairs. The numerical results confirm that the modified upper bound in \cref{eq:modifiedCH} is respected across the entire $\vartheta$ range.

To investigate higher-order correlations in $J/\psi$ decay processes, let us consider the measurement settings defined by the directions:
\begin{align}
\label{eq:measurementSetting2}
\boldsymbol{n}_p &= (0,0,1),\quad \boldsymbol{n}_p' = (0,1,0),\notag \\
\boldsymbol{n}_{\bar{p}} &= \frac{1}{\sqrt{2}}(0,1,-1),\quad 
\boldsymbol{n}_{\bar{p}}^{'} = \frac{1}{\sqrt{2}}(0,1,1)\ .
\end{align}
Accordingly, this measurement setting gives the magnitude of third-order cumulant for the state in \cref{eq:Xstate}:
\begin{align}
\label{eq:k3_Xstate}
|\kappa_3^{(\psi)}| = \frac{\sqrt{2}\alpha_Y^6}{16}|(t_2-t_3)((t_2-t_3)^2 - 3t_1- 1 )| ,
\end{align}

The third-order cumulants $\kappa_3^{(\psi)}$ for entangled $\Sigma^+\bar{\Sigma}^-$ and $\Lambda\bar{\Lambda}$ pairs are presented in \cref{fig:sigmaLambdaSkew} as functions of $\vartheta$.  Evidently, $\kappa_3^{(\psi)}$ violates the classical upper bound derived in \cref{theorem:CHcumulant} within certain scattering angle ranges, exhibiting higher-order quantum correlations. However, when accounting for the timelike modification, neither channel violates the stricter bound from \cref{corollary:hyperonCHcumulant}, as depicted in \cref{fig:sigmaLambdaSkew}.

Crucially, comparing \cref{fig:sigmaLambdaMeanNoviolation,fig:sigmaLambdaSkew} demonstrates that the violation windows (in terms of $\vartheta$) for both $\Sigma^+\bar{\Sigma}^-$ and $\Lambda\bar{\Lambda}$ pairs are wider for the third-order cumulant in \cref{theorem:CHcumulant} than for the CH inequality in \cref{eq:CHhyperon}. This expanded range suggests that the third-order cumulant, as a higher-order statistical measure, is more sensitive to the non-linear quantum correlations.


\subsection{The fourth-order central moment}
While the first three cumulants correspond directly to moments or central moments (representing the mean, variance, and skewness, respectively), higher-order cumulants ($n \ge 4$) define a more complex algebraic structure involving polynomial combinations of moments. The fourth-order cumulant is the first instance where this distinction becomes significant.
Despite their theoretical importance, deriving tight bounds for higher-order cumulants remains a nontrivial mathematical challenge \cite{prohorov69,dubkov76,zhang25}.

Instead, we investigate the fourth-order central moment rather than the cumulant, which may provide useful insights of contextuality. Based on the constraints of bounded random variables \cite{egozcue12}, we have:
\begin{corollary}
\label{corollary:hyperonCHfourthCentr}
For any bipartite system, local realism imposes the following constraint on the fourth-order central moment of the CH operator:
\begin{align}
\label{eq:idealfourthCentrbound}
\mu_4(\mathcal{B}_{\mathrm{CH}}) \leq & \frac{ \alpha^8}{12}\ .
\end{align}
Accounting for the fraction $\beta$ of spacelike separated events, the modified upper bound becomes:
\begin{align}
\label{eq:modifiedfourthCentrbound}
\mu_4(\mathcal{B}_{\text{CH}}) \leq & \frac{(2-\beta)^4 \alpha^8}{12}\ .
\end{align}
\end{corollary}


Adopting the measurement settings defined in \cref{eq:measurementSetting1},
the analytical expression for the fourth-order central moment of $\mathcal{B}_{\mathrm{CH}}$ for $\chi_{c0}/\eta_c \to \Lambda\bar{\Lambda}$ is derived as:
\begin{align}
\mu_4^{(\chi/\eta)}= \frac{\alpha_\Lambda^8}{16} \left[ 2 - 2\sin(2\phi) + 3\cos^2(2\phi) \right]\ ,
\end{align}
which achieves its maximum value when $\sin(2\phi) = -\frac{1}{3}$.
As illustrated in \cref{fig:etacChic0FourthCentral}, 
the quantum prediction for $\mu_4$ significantly exceeds both the classical limit in \cref{eq:idealfourthCentrbound} and the stricter timelike-modified bound in \cref{eq:modifiedfourthCentrbound} for both $\chi_{c0}$ and $\eta_c$ channels.

Turning to the entangled hyperon-antihyperon pairs produced in $J/\psi$ decay, and using the measurement settings in \cref{eq:measurementSetting2}, the fourth-order central moment of $\mathcal{B}_{\mathrm{CH}}$ is given by 
\begin{align}
\mu_4^{(\psi)}= \frac{\alpha_\Lambda^8}{64} \left[ -3(t_2 - t_3)^4 + 4(3t_1 - 1)(t_2 - t_3)^2 + 8(1 + t_1) \right]\ .
\end{align}
The $\mu_4^{(\psi)}$ for $\Lambda\bar{\Lambda}$ and $\Sigma^+\bar{\Sigma}^-$ pairs are numerically evaluated and depicted in \cref{fig:JpsiLsFourthCentral} as functions of the scattering angle $\vartheta$. Remarkably, it is observed that $\mu_4$ violates the classical upper bound in \cref{eq:idealfourthCentrbound} across the entire range of scattering angles $\vartheta$.

For simplicity, let us consider the CHSH operator composed with the orthogonal measurement configuration ($\boldsymbol{n}_p\perp\boldsymbol{n}_p^\prime$, $\boldsymbol{n}_{\bar{p}}\perp\boldsymbol{n}_{\bar{p}}^\prime$), i.e.,
\begin{align}
\mathcal{B}_{\mathrm{CHSH}} = \boldsymbol{n}_p\cdot \boldsymbol{\sigma} \otimes (\boldsymbol{n}_{\bar{p}} + \boldsymbol{n}_{\bar{p}}^\prime)\cdot \boldsymbol{\sigma} + \boldsymbol{n}_p^\prime\cdot \boldsymbol{\sigma} \otimes (\boldsymbol{n}_{\bar{p}} - \boldsymbol{n}_{\bar{p}}^\prime)\cdot \boldsymbol{\sigma}\ .
\end{align}
In the fourth-order central moment of $\mathcal{B}_{\mathrm{CHSH}}$, the leading term is given by
\begin{align}
\mathcal{B}_{\mathrm{CHSH}}^4 =& 16 I \otimes I + [\boldsymbol{n}_p\cdot\boldsymbol{\sigma},\boldsymbol{n}_p^\prime\cdot\boldsymbol{\sigma}]^2 \otimes [\boldsymbol{n}_{\bar{p}}\cdot\boldsymbol{\sigma},\boldsymbol{n}_{\bar{p}}^\prime\cdot\boldsymbol{\sigma}]^2 \notag\\
&\quad + 8[\boldsymbol{n}_p\cdot\boldsymbol{\sigma},\boldsymbol{n}_p^\prime\cdot\boldsymbol{\sigma}] \otimes [\boldsymbol{n}_{\bar{p}}\cdot\boldsymbol{\sigma},\boldsymbol{n}_{\bar{p}}^\prime\cdot\boldsymbol{\sigma}]\ .
\end{align}
Recent studies suggest that non-trivial expectation values of squared commutators serve as signatures of state-independent contextuality \cite{yang23}, which may explain the full-range violation of the ideal upper bound in \cref{eq:idealfourthCentrbound} by the fourth-order central moment for the entangled hyperon-antihyperon pairs produced in $J/\psi$ decay as shown in \cref{fig:JpsiLsFourthCentral}. To clarify the exact role of the fourth-order central moment in revealing contextuality, further investigations are required, which we leave for future work.


\subsection{Dynamical Origin of Correlations: Electromagnetic Form Factors}
\label{sec:EMFF}

In the preceding sections, the parameters characterizing the quantum state ($\alpha_{\psi}, \Delta\Phi$) and the kinematic threshold ($\beta$) were treated as phenomenological inputs derived from experiment. To fundamentally understand the conditions required to observe high-order nonlocality violations, we must examine the interplay between the underlying QCD dynamics and the kinematic regime of the collider.

From \cref{eq:CHmax_at_pi2,eq:k3_Xstate}, it is evident that the potential for violating local realism through both the CH inequality and high-order cumulants is intrinsically linked to the eigenvalues of the correlation matrix $T$, which are functions of the decay parameters $\alpha_\psi$ and $\Delta\Phi$.
For the process $e^{+}e^{-}\rightarrow J/\psi\rightarrow Y\overline{Y}$, the spin correlations are dictated by the interference between the magnetic ($G_{M}^{\psi}$) and electric ($G_{E}^{\psi}$) form factors. The decay parameter $\alpha_{\psi}$, which determines the weight of the entangled components in the density matrix, is related to the form factors by \cite{faldt17}:
\begin{align}
\alpha_\psi = \frac{s|G_M^\psi|^2 - 4M^2|G_E^\psi|^2}{s|G_M^\psi|^2 + 4M^2|G_E^\psi|^2}\ ,
\end{align} 
where $\sqrt{s}$ is the center-of-mass energy and $M$ is the baryon mass. This relationship maps the continuous parameter space of form factors onto the observable correlation strength. 

Two extreme dynamical regimes yield maximal entanglement, ideal for testing local realism. In the magnetic dominance limit ($|G_{E}^{\psi}|\rightarrow 0$), $\alpha_{\psi}\rightarrow 1$, preserving helicity. Conversely, in the electric dominance limit ($|G_{M}^{\psi}|\rightarrow 0$), $\alpha_{\psi}\rightarrow -1$. Both extremes correspond to maximally entangled spin states. However, experimental data (see \cref{tab:combined_params}) indicates that many hyperon channels, such as $J/\psi\rightarrow\Sigma^{+}\overline{\Sigma}^{-}$, lie in a ``mixed" regime where $\alpha_{\psi}\approx -0.508$. This interference between electric and magnetic amplitudes induces a deviation from the maximally entangled state. Furthermore, the relative phase $\Delta\Phi$ generates a transverse polarization $P_y$. This introduces single-particle terms into the density matrix that deviate from the Bell diagonal state structure, thereby explicitly reducing the entanglement concurrence below the limit set by $\alpha_{\psi}$ alone \cite{wu24}.


\section{Conclusions}
In summary, addressing the unexplored regime beyond linear statistical moments, the higher-order quantum correlations are investigated in entangled hyperon-antihyperon systems via statistical cumulants of the generalized Clauser-Horne operator. A generalized CH inequality based on the second-order cumulant is formulated, and further extended to derive a rigorous third-order skewness bound.

The numerical analysis reveals distinct violation patterns. Specifically, the $\chi_{c0} \to \Lambda\bar{\Lambda}$ process exhibits a pronounced violation of the skewness constraint, surpassing the classical limit by a clear margin. Notably, this violation is shown to be robust against the kinematic contamination from timelike separated events, confirming that the higher-order quantum correlation in this channel is observable in high-energy experiments such as BESIII and Belle II. Crucially, for the spin-singlet states in $\eta_c$ and $\chi_{c0}$ decays, the optimal kinematic configurations maximizing this third-order violation differ from those of the generalized CH inequality, demonstrating that higher-order statistics offer complementary sensitivity to quantum correlations.

Furthermore, the $J/\psi \to \Sigma^+\bar{\Sigma}^-$ and $\Lambda\bar{\Lambda}$ processes violate the third-order cumulant criterion with a strictly wider window in terms of the scattering angle compared to the generalized CH inequality. However, when incorporating the timelike modification, neither channel violates the stricter generalized CH inequality or the modified third-order bound. Moreover, the exploration of the fourth-order central moment shows that it exceeds the classical limit across the entire scattering angle range for hyperon pairs from $J/\psi$ decay, suggesting a potential link to state-independent contextuality.

Finally, the dynamical origins of these correlations are analyzed in terms of electromagnetic form factors. It is found that the interference between electric and magnetic form factors leads to a departure from maximal entanglement. This insight underscores the intricate interplay between QCD dynamics and quantum correlations in hadronic systems.

\section*{Acknowledgements}
This work was supported in part by National Natural Science Foundation of China (NSFC) under the Grants 12475087 and 12235008, and University of Chinese Academy of Sciences.

\appendix 
\section{Proof of Lemma \ref{lemma:CH}}
\label{app:ProofCH}
The generalized CH operator is defined as 
\begin{align}
\mathcal{B}_{\text{CH}} 
& \equiv E_+(\boldsymbol{a}) \otimes E_+(\boldsymbol{b})-E_+\left(\boldsymbol{a}\right) \otimes E_+\left(\boldsymbol{b}^{\prime}\right)\notag\\
& \quad +E_+\left(\boldsymbol{a}^{\prime}\right) \otimes E_+(\boldsymbol{b}) +E_+\left(\boldsymbol{a}^{\prime}\right) \otimes E_+\left(\boldsymbol{b}^{\prime}\right)\notag\\
& \quad -\left(1+ \eta_b\right) E_+\left(\boldsymbol{a}^{\prime}\right) \otimes I -\left(1+ \eta_a\right) I \otimes E_+(\boldsymbol{b})\notag\\
& \quad +\frac{\left(1+ \eta_a\right)\left(1+ \eta_b\right)-\left|\alpha_a\alpha_b\right|}{2} I\otimes I\ .
\end{align}
Here, the general POVM operator $E_+(\boldsymbol{n})$ is given by
\begin{align}
E_+(\boldsymbol{n})=\frac{1}{2}(\xi+\alpha \boldsymbol{n}\cdot\boldsymbol{\sigma})\ ,
\end{align}
with $\boldsymbol{\sigma}$ is the vector of Pauli matrices.

The second-order cumulant of $\mathcal{B}_{\mathrm{CH}}$ is defined as follows
\begin{align}
\kappa_2(\mathcal{B}_{\mathrm{CH}})
=\langle \mathcal{B}_{\mathrm{CH}}^2\rangle - \langle \mathcal{B}_{\mathrm{CH}}\rangle^2\ .
\end{align}
First, it can be derived that the square of the CH operator defined in \cref{def:CHoperator} satisfies the identity:
\begin{align}
\mathcal{B}_{\text{CH}}^2=-|\alpha_a\alpha_b| \mathcal{B}_{\text{CH}}-\alpha_a^2\alpha_b^2\mathcal{C}\ ,
\end{align}
where $\mathcal{C}=[\boldsymbol{a}\cdot\boldsymbol{\sigma},\boldsymbol{a}'\cdot\boldsymbol{\sigma}]\otimes [\boldsymbol{b}\cdot\boldsymbol{\sigma},\boldsymbol{b}'\cdot\boldsymbol{\sigma}]/16$.
Thus, the nonnegativity of the second-order cumulant of $\mathcal{B}_{\mathrm{CH}}$ implies that
\begin{align}
-|\alpha_a\alpha_b| \langle\mathcal{B}_{\mathrm{CH}}\rangle-\langle\mathcal{B}_{\mathrm{CH}}\rangle^2-\alpha_a^2\alpha_b^2\langle\mathcal{C}\rangle \geq 0\ ,
\end{align}
In any LHV theory, there exists a joint distribution model for the four dichotomic observables and all local observables commute \cite{fine82,landau87}, i.e., $\langle\mathcal{C}\rangle=0$, which leads to
\begin{align}
- |\alpha_a\alpha_b| \langle\mathcal{B}_{\mathrm{CH}}\rangle_{\mathrm{lhv}}-\langle\mathcal{B}_{\mathrm{CH}}\rangle_{\mathrm{lhv}}^2 \geq 0\ .
\end{align}
By solving this inequality, one can obtain that $\langle\mathcal{B}_{\mathrm{CH}}\rangle_{\mathrm{lhv}}$ is strictly bounded in the interval:
\begin{align}
\label{eq:LHV_domain}
\langle\mathcal{B}_{\mathrm{CH}}\rangle_{\mathrm{lhv}} \in \left[ -|\alpha_a\alpha_b|, \, 0 \right]\ .
\end{align}

\section{Proof of the upper bounds on high-order central moments}
\label{app:ProofCumulantBounds}
For any random variable $X$ supported on a bounded interval $[A, B]$ of length $L = B - A$, the statistical cumulants are strictly constrained by the geometry of the support. The third-order cumulant (skewness), $\kappa_3 = \langle (X - \mu)^3 \rangle$, satisfies the algebraic inequality for bounded variables \cite{egozcue12}:
\begin{align}
|\kappa_3| \le \frac{(B-A)^3}{8}.
\end{align}
Substituting the support length $L = |\alpha_a \alpha_b|$ derived from the LHV constraint, we obtain the necessary condition for local realism:
\begin{align}
|\kappa_3(\mathcal{B}_{CH})| \le \frac{|\alpha_a \alpha_b|^3}{8}.
\end{align}
This completes the proof of \cref{theorem:CHcumulant}. The proof of \cref{corollary:hyperonCHcumulant} can be similarly established.

Similarly, for the fourth-order central moment $\mu_4 = \langle (X - \mu)^4 \rangle$, the upper bound for a random variable defined on an interval of length $L$ is given by limits \cite{egozcue12}:
\begin{align}
\mu_4 \le \frac{L^4}{12}.
\end{align}
Substituting $L = |\alpha_a \alpha_b|$ (and setting $|\alpha_a|=|\alpha_b|=\alpha$ for simplicity), we arrive at the bound:
\begin{align}
\mu_4(\mathcal{B}_{CH}) \le \frac{\alpha^8}{12}.
\end{align}
This completes the proof of \cref{corollary:hyperonCHfourthCentr}.


\begin{thebibliography}{52}%
	\makeatletter
	\providecommand \@ifxundefined [1]{%
		\@ifx{#1\undefined}
	}%
	\providecommand \@ifnum [1]{%
		\ifnum #1\expandafter \@firstoftwo
		\else \expandafter \@secondoftwo
		\fi
	}%
	\providecommand \@ifx [1]{%
		\ifx #1\expandafter \@firstoftwo
		\else \expandafter \@secondoftwo
		\fi
	}%
	\providecommand \natexlab [1]{#1}%
	\providecommand \enquote  [1]{``#1''}%
	\providecommand \bibnamefont  [1]{#1}%
	\providecommand \bibfnamefont [1]{#1}%
	\providecommand \citenamefont [1]{#1}%
	\providecommand \href@noop [0]{\@secondoftwo}%
	\providecommand \href [0]{\begingroup \@sanitize@url \@href}%
	\providecommand \@href[1]{\@@startlink{#1}\@@href}%
	\providecommand \@@href[1]{\endgroup#1\@@endlink}%
	\providecommand \@sanitize@url [0]{\catcode `\\12\catcode `\$12\catcode `\&12\catcode `\#12\catcode `\^12\catcode `\_12\catcode `\%12\relax}%
	\providecommand \@@startlink[1]{}%
	\providecommand \@@endlink[0]{}%
	\providecommand \url  [0]{\begingroup\@sanitize@url \@url }%
	\providecommand \@url [1]{\endgroup\@href {#1}{\urlprefix }}%
	\providecommand \urlprefix  [0]{URL }%
	\providecommand \Eprint [0]{\href }%
	\providecommand \doibase [0]{http://dx.doi.org/}%
	\providecommand \selectlanguage [0]{\@gobble}%
	\providecommand \bibinfo  [0]{\@secondoftwo}%
	\providecommand \bibfield  [0]{\@secondoftwo}%
	\providecommand \translation [1]{[#1]}%
	\providecommand \BibitemOpen [0]{}%
	\providecommand \bibitemStop [0]{}%
	\providecommand \bibitemNoStop [0]{.\EOS\space}%
	\providecommand \EOS [0]{\spacefactor3000\relax}%
	\providecommand \BibitemShut  [1]{\csname bibitem#1\endcsname}%
	\let\auto@bib@innerbib\@empty
	\bibitem [{\citenamefont {Einstein}\ \emph {et~al.}(1935)\citenamefont {Einstein} \emph {et~al.}}]{einstein35}%
	\BibitemOpen
	\bibfield  {author} {\bibinfo {author} {\bibfnamefont {A.}~\bibnamefont {Einstein}} \emph {et~al.},\ }\bibfield  {title} {\emph {\bibinfo {title} {Can quantum-mechanical description of physical reality be considered complete?}\ }}\href {\doibase 10.1103/PhysRev.47.777} {\bibfield  {journal} {\bibinfo  {journal} {Physical Review}\ }\textbf {\bibinfo {volume} {47}},\ \bibinfo {pages} {777} (\bibinfo {year} {1935})}\BibitemShut {NoStop}%
	\bibitem [{\citenamefont {Bell}(1964)}]{bell64}%
	\BibitemOpen
	\bibfield  {author} {\bibinfo {author} {\bibfnamefont {J.~S.}\ \bibnamefont {Bell}},\ }\bibfield  {title} {\emph {\bibinfo {title} {On the {Einstein Podolsky Rosen} paradox},\ }}\href {\doibase 10.1103/PhysicsPhysiqueFizika.1.195} {\bibfield  {journal} {\bibinfo  {journal} {Physics Physique Fizika}\ }\textbf {\bibinfo {volume} {1}},\ \bibinfo {pages} {195–200} (\bibinfo {year} {1964})}\BibitemShut {NoStop}%
	\bibitem [{\citenamefont {Clauser}\ \emph {et~al.}(1969)\citenamefont {Clauser} \emph {et~al.}}]{clauser69}%
	\BibitemOpen
	\bibfield  {author} {\bibinfo {author} {\bibfnamefont {J.~F.}\ \bibnamefont {Clauser}} \emph {et~al.},\ }\bibfield  {title} {\emph {\bibinfo {title} {Proposed experiment to test local hidden-variable theories},\ }}\href {\doibase 10.1103/PhysRevLett.23.880} {\bibfield  {journal} {\bibinfo  {journal} {Physical Review Letters}\ }\textbf {\bibinfo {volume} {23}},\ \bibinfo {pages} {880–884} (\bibinfo {year} {1969})}\BibitemShut {NoStop}%
	\bibitem [{\citenamefont {Freedman}\ and\ \citenamefont {Clauser}(1972)}]{freedman72}%
	\BibitemOpen
	\bibfield  {author} {\bibinfo {author} {\bibfnamefont {S.~J.}\ \bibnamefont {Freedman}}\ and\ \bibinfo {author} {\bibfnamefont {J.~F.}\ \bibnamefont {Clauser}},\ }\bibfield  {title} {\emph {\bibinfo {title} {Experimental test of local hidden-variable theories},\ }}\href {\doibase 10.1103/PhysRevLett.28.938} {\bibfield  {journal} {\bibinfo  {journal} {Physical Review Letters}\ }\textbf {\bibinfo {volume} {28}},\ \bibinfo {pages} {938–941} (\bibinfo {year} {1972})}\BibitemShut {NoStop}%
	\bibitem [{\citenamefont {Clauser}\ and\ \citenamefont {Horne}(1974)}]{clauser74e}%
	\BibitemOpen
	\bibfield  {author} {\bibinfo {author} {\bibfnamefont {J.~F.}\ \bibnamefont {Clauser}}\ and\ \bibinfo {author} {\bibfnamefont {M.~A.}\ \bibnamefont {Horne}},\ }\bibfield  {title} {\emph {\bibinfo {title} {Experimental consequences of objective local theories},\ }}\href {\doibase 10.1103/PhysRevD.10.526} {\bibfield  {journal} {\bibinfo  {journal} {Physical Review D}\ }\textbf {\bibinfo {volume} {10}},\ \bibinfo {pages} {526–535} (\bibinfo {year} {1974})}\BibitemShut {NoStop}%
	\bibitem [{\citenamefont {Aspect}\ \emph {et~al.}(1982)\citenamefont {Aspect} \emph {et~al.}}]{aspect82}%
	\BibitemOpen
	\bibfield  {author} {\bibinfo {author} {\bibfnamefont {A.}~\bibnamefont {Aspect}} \emph {et~al.},\ }\bibfield  {title} {\emph {\bibinfo {title} {Experimental test of {Bell's} inequalities using time-varying analyzers},\ }}\href {\doibase 10.1103/PhysRevLett.49.1804} {\bibfield  {journal} {\bibinfo  {journal} {Physical Review Letters}\ }\textbf {\bibinfo {volume} {49}},\ \bibinfo {pages} {1804–1807} (\bibinfo {year} {1982})}\BibitemShut {NoStop}%
	\bibitem [{\citenamefont {Wiseman}\ \emph {et~al.}(2007)\citenamefont {Wiseman} \emph {et~al.}}]{wiseman07}%
	\BibitemOpen
	\bibfield  {author} {\bibinfo {author} {\bibfnamefont {H.~M.}\ \bibnamefont {Wiseman}} \emph {et~al.},\ }\bibfield  {title} {\emph {\bibinfo {title} {Steering, entanglement, nonlocality, and the {Einstein-Podolsky-Rosen} paradox},\ }}\href {\doibase 10.1103/PhysRevLett.98.140402} {\bibfield  {journal} {\bibinfo  {journal} {Phys Rev Lett}\ }\textbf {\bibinfo {volume} {98}},\ \bibinfo {pages} {140402} (\bibinfo {year} {2007})}\BibitemShut {NoStop}%
	\bibitem [{\citenamefont {Cavalcanti}\ \emph {et~al.}(2009)\citenamefont {Cavalcanti} \emph {et~al.}}]{cavalcanti09}%
	\BibitemOpen
	\bibfield  {author} {\bibinfo {author} {\bibfnamefont {E.~G.}\ \bibnamefont {Cavalcanti}} \emph {et~al.},\ }\bibfield  {title} {\emph {\bibinfo {title} {Experimental criteria for steering and the {Einstein-Podolsky-Rosen} paradox},\ }}\href {\doibase 10.1103/PhysRevA.80.032112} {\bibfield  {journal} {\bibinfo  {journal} {Physical Review A}\ }\textbf {\bibinfo {volume} {80}},\ \bibinfo {pages} {032112} (\bibinfo {year} {2009})}\BibitemShut {NoStop}%
	\bibitem [{\citenamefont {Gühne}\ and\ \citenamefont {Tóth}(2009)}]{OG09E}%
	\BibitemOpen
	\bibfield  {author} {\bibinfo {author} {\bibfnamefont {O.}~\bibnamefont {Gühne}}\ and\ \bibinfo {author} {\bibfnamefont {G.}~\bibnamefont {Tóth}},\ }\bibfield  {title} {\emph {\bibinfo {title} {Entanglement detection},\ }}\href {\doibase 10.1016/j.physrep.2009.02.004} {\bibfield  {journal} {\bibinfo  {journal} {Physics Reports}\ }\textbf {\bibinfo {volume} {474}},\ \bibinfo {pages} {1–75} (\bibinfo {year} {2009})}\BibitemShut {NoStop}%
	\bibitem [{\citenamefont {Uola}\ \emph {et~al.}(2020)\citenamefont {Uola} \emph {et~al.}}]{UR20Q}%
	\BibitemOpen
	\bibfield  {author} {\bibinfo {author} {\bibfnamefont {R.}~\bibnamefont {Uola}} \emph {et~al.},\ }\bibfield  {title} {\emph {\bibinfo {title} {Quantum steering},\ }}\href {\doibase 10.1103/RevModPhys.92.015001} {\bibfield  {journal} {\bibinfo  {journal} {Reviews of Modern Physics}\ }\textbf {\bibinfo {volume} {92}} (\bibinfo {year} {2020}),\ 10.1103/RevModPhys.92.015001}\BibitemShut {NoStop}%
	\bibitem [{\citenamefont {Brunner}\ \emph {et~al.}(2014)\citenamefont {Brunner} \emph {et~al.}}]{brunner14}%
	\BibitemOpen
	\bibfield  {author} {\bibinfo {author} {\bibfnamefont {N.}~\bibnamefont {Brunner}} \emph {et~al.},\ }\bibfield  {title} {\emph {\bibinfo {title} {Bell nonlocality},\ }}\href {\doibase 10.1103/RevModPhys.86.419} {\bibfield  {journal} {\bibinfo  {journal} {Reviews of Modern Physics}\ }\textbf {\bibinfo {volume} {86}},\ \bibinfo {pages} {419–478} (\bibinfo {year} {2014})}\BibitemShut {NoStop}%
	\bibitem [{\citenamefont {Branciard}\ \emph {et~al.}(2007)\citenamefont {Branciard} \emph {et~al.}}]{branciard07}%
	\BibitemOpen
	\bibfield  {author} {\bibinfo {author} {\bibfnamefont {C.}~\bibnamefont {Branciard}} \emph {et~al.},\ }\bibfield  {title} {\emph {\bibinfo {title} {Experimental falsification of {Leggett's} nonlocal variable model},\ }}\href {\doibase 10.1103/PhysRevLett.99.210407} {\bibfield  {journal} {\bibinfo  {journal} {Physical Review Letters}\ }\textbf {\bibinfo {volume} {99}},\ \bibinfo {pages} {210407} (\bibinfo {year} {2007})}\BibitemShut {NoStop}%
	\bibitem [{\citenamefont {Paterek}\ \emph {et~al.}(2007)\citenamefont {Paterek} \emph {et~al.}}]{paterek07}%
	\BibitemOpen
	\bibfield  {author} {\bibinfo {author} {\bibfnamefont {T.}~\bibnamefont {Paterek}} \emph {et~al.},\ }\bibfield  {title} {\emph {\bibinfo {title} {Experimental test of nonlocal realistic theories without the rotational symmetry assumption},\ }}\href {\doibase 10.1103/PhysRevLett.99.210406} {\bibfield  {journal} {\bibinfo  {journal} {Physical Review Letters}\ }\textbf {\bibinfo {volume} {99}},\ \bibinfo {pages} {210406} (\bibinfo {year} {2007})}\BibitemShut {NoStop}%
	\bibitem [{\citenamefont {Gr{\"o}blacher}\ \emph {et~al.}(2007)\citenamefont {Gr{\"o}blacher} \emph {et~al.}}]{groblacher07}%
	\BibitemOpen
	\bibfield  {author} {\bibinfo {author} {\bibfnamefont {S.}~\bibnamefont {Gr{\"o}blacher}} \emph {et~al.},\ }\bibfield  {title} {\emph {\bibinfo {title} {An experimental test of non-local realism},\ }}\href {\doibase 10.1038/nature05677} {\bibfield  {journal} {\bibinfo  {journal} {Nature}\ }\textbf {\bibinfo {volume} {446}},\ \bibinfo {pages} {871–875} (\bibinfo {year} {2007})}\BibitemShut {NoStop}%
	\bibitem [{\citenamefont {Branciard}\ \emph {et~al.}(2008)\citenamefont {Branciard} \emph {et~al.}}]{branciard08}%
	\BibitemOpen
	\bibfield  {author} {\bibinfo {author} {\bibfnamefont {C.}~\bibnamefont {Branciard}} \emph {et~al.},\ }\bibfield  {title} {\emph {\bibinfo {title} {Testing quantum correlations versus single-particle properties within {Leggett's} model and beyond},\ }}\href {\doibase 10.1038/nphys1020} {\bibfield  {journal} {\bibinfo  {journal} {Nature Physics}\ }\textbf {\bibinfo {volume} {4}},\ \bibinfo {pages} {681–685} (\bibinfo {year} {2008})}\BibitemShut {NoStop}%
	\bibitem [{\citenamefont {Budroni}\ \emph {et~al.}(2022)\citenamefont {Budroni} \emph {et~al.}}]{budroni22}%
	\BibitemOpen
	\bibfield  {author} {\bibinfo {author} {\bibfnamefont {C.}~\bibnamefont {Budroni}} \emph {et~al.},\ }\bibfield  {title} {\emph {\bibinfo {title} {{Kochen-Specker} contextuality},\ }}\href {\doibase 10.1103/revmodphys.94.045007} {\bibfield  {journal} {\bibinfo  {journal} {Reviews of Modern Physics}\ }\textbf {\bibinfo {volume} {94}},\ \bibinfo {pages} {045007} (\bibinfo {year} {2022})}\BibitemShut {NoStop}%
	\bibitem [{\citenamefont {T{\" o}rnqvist}(1981)}]{tornqvist81}%
	\BibitemOpen
	\bibfield  {author} {\bibinfo {author} {\bibfnamefont {N.~A.}\ \bibnamefont {T{\" o}rnqvist}},\ }\bibfield  {title} {\emph {\bibinfo {title} {Suggestion for einstein-podolsky-rosen experiments using reactions like $e^ + e^ - \to \lambda \bar \lambda \to \pi ^ - p\pi ^ + \bar p$},\ }}\href {\doibase 10.1007/BF00715204} {\bibfield  {journal} {\bibinfo  {journal} {Foundations of Physics}\ }\textbf {\bibinfo {volume} {11}},\ \bibinfo {pages} {171–177} (\bibinfo {year} {1981})}\BibitemShut {NoStop}%
	\bibitem [{\citenamefont {Li}\ and\ \citenamefont {Qiao}(2006)}]{li06}%
	\BibitemOpen
	\bibfield  {author} {\bibinfo {author} {\bibfnamefont {J.}~\bibnamefont {Li}}\ and\ \bibinfo {author} {\bibfnamefont {C.-F.}\ \bibnamefont {Qiao}},\ }\bibfield  {title} {\emph {\bibinfo {title} {Feasibility of testing local hidden variable theories in a charm factory},\ }}\href {\doibase 10.1103/PhysRevD.74.076003} {\bibfield  {journal} {\bibinfo  {journal} {Physical Review D}\ }\textbf {\bibinfo {volume} {74}},\ \bibinfo {pages} {076003} (\bibinfo {year} {2006})}\BibitemShut {NoStop}%
	\bibitem [{\citenamefont {Li}\ and\ \citenamefont {Qiao}(2009)}]{li09n}%
	\BibitemOpen
	\bibfield  {author} {\bibinfo {author} {\bibfnamefont {J.}~\bibnamefont {Li}}\ and\ \bibinfo {author} {\bibfnamefont {C.-F.}\ \bibnamefont {Qiao}},\ }\bibfield  {title} {\emph {\bibinfo {title} {New possibilities for testing local realism in high energy physics},\ }}\href {\doibase https://doi.org/10.1016/j.physleta.2009.09.057} {\bibfield  {journal} {\bibinfo  {journal} {Physics Letters A}\ }\textbf {\bibinfo {volume} {373}},\ \bibinfo {pages} {4311–4314} (\bibinfo {year} {2009})}\BibitemShut {NoStop}%
	\bibitem [{\citenamefont {Li}\ and\ \citenamefont {Qiao}(2010)}]{li10}%
	\BibitemOpen
	\bibfield  {author} {\bibinfo {author} {\bibfnamefont {J.}~\bibnamefont {Li}}\ and\ \bibinfo {author} {\bibfnamefont {C.}~\bibnamefont {Qiao}},\ }\bibfield  {title} {\emph {\bibinfo {title} {Testing local realism in {$P \to VV$} decays},\ }}\href {\doibase 10.1007/s11433-010-0202-2} {\bibfield  {journal} {\bibinfo  {journal} {Science China Physics, Mechanics and Astronomy}\ }\textbf {\bibinfo {volume} {53}},\ \bibinfo {pages} {870–875} (\bibinfo {year} {2010})}\BibitemShut {NoStop}%
	\bibitem [{\citenamefont {Shi}\ and\ \citenamefont {Yang}(2020)}]{shi20}%
	\BibitemOpen
	\bibfield  {author} {\bibinfo {author} {\bibfnamefont {Y.}~\bibnamefont {Shi}}\ and\ \bibinfo {author} {\bibfnamefont {J.-C.}\ \bibnamefont {Yang}},\ }\bibfield  {title} {\emph {\bibinfo {title} {Entangled baryons: Violation of inequalities based on local realism assuming dependence of decays on hidden variables},\ }}\href {\doibase 10.1140/epjc/s10052-020-7684-5} {\bibfield  {journal} {\bibinfo  {journal} {The European Physical Journal C}\ }\textbf {\bibinfo {volume} {80}},\ \bibinfo {pages} {116} (\bibinfo {year} {2020})}\BibitemShut {NoStop}%
	\bibitem [{\citenamefont {Fabbrichesi}\ \emph {et~al.}(2021)\citenamefont {Fabbrichesi}, \citenamefont {Floreanini},\ and\ \citenamefont {Panizzo}}]{fabbrichesi21}%
	\BibitemOpen
	\bibfield  {author} {\bibinfo {author} {\bibfnamefont {M.}~\bibnamefont {Fabbrichesi}}, \bibinfo {author} {\bibfnamefont {R.}~\bibnamefont {Floreanini}}, \ and\ \bibinfo {author} {\bibfnamefont {G.}~\bibnamefont {Panizzo}},\ }\bibfield  {title} {\emph {\bibinfo {title} {Testing {Bell} inequalities at the lhc with top-quark pairs},\ }}\href {\doibase 10.1103/PhysRevLett.127.161801} {\bibfield  {journal} {\bibinfo  {journal} {Physical Review Letters}\ }\textbf {\bibinfo {volume} {127}},\ \bibinfo {pages} {161801} (\bibinfo {year} {2021})}\BibitemShut {NoStop}%
	\bibitem [{\citenamefont {Afik}\ and\ \citenamefont {de~Nova}(2023)}]{afik23}%
	\BibitemOpen
	\bibfield  {author} {\bibinfo {author} {\bibfnamefont {Y.}~\bibnamefont {Afik}}\ and\ \bibinfo {author} {\bibfnamefont {J.~R.~M.}\ \bibnamefont {de~Nova}},\ }\bibfield  {title} {\emph {\bibinfo {title} {Quantum discord and steering in top quarks at the lhc},\ }}\href {\doibase 10.1103/PhysRevLett.130.221801} {\bibfield  {journal} {\bibinfo  {journal} {Physical Review Letters}\ }\textbf {\bibinfo {volume} {130}},\ \bibinfo {pages} {221801} (\bibinfo {year} {2023})}\BibitemShut {NoStop}%
	\bibitem [{\citenamefont {Bertlmann}\ \emph {et~al.}(2004)\citenamefont {Bertlmann} \emph {et~al.}}]{bertlmann04}%
	\BibitemOpen
	\bibfield  {author} {\bibinfo {author} {\bibfnamefont {R.~A.}\ \bibnamefont {Bertlmann}} \emph {et~al.},\ }\bibfield  {title} {\emph {\bibinfo {title} {Violation of a {Bell} inequality in particle physics experimentally verified?}\ }}\href {\doibase https://doi.org/10.1016/j.physleta.2004.10.006} {\bibfield  {journal} {\bibinfo  {journal} {Physics Letters A}\ }\textbf {\bibinfo {volume} {332}},\ \bibinfo {pages} {355–360} (\bibinfo {year} {2004})}\BibitemShut {NoStop}%
	\bibitem [{\citenamefont {Afriat}\ and\ \citenamefont {Selleri}(2013)}]{afriat13}%
	\BibitemOpen
	\bibfield  {author} {\bibinfo {author} {\bibfnamefont {A.}~\bibnamefont {Afriat}}\ and\ \bibinfo {author} {\bibfnamefont {F.}~\bibnamefont {Selleri}},\ }\href {\doibase 10.1007/978-1-4899-0254-2} {\emph {\bibinfo {title} {The {Einstein, Podolsky, and Rosen} Paradox in Atomic, Nuclear, and Particle Physics}}},\ \bibinfo {edition} {1st}\ ed.\ (\bibinfo  {publisher} {Springer New York, NY},\ \bibinfo {year} {2013})\ p.\ \bibinfo {pages} {248}\BibitemShut {NoStop}%
	\bibitem [{\citenamefont {Hiesmayr}(2015)}]{hiesmayr15}%
	\BibitemOpen
	\bibfield  {author} {\bibinfo {author} {\bibfnamefont {B.~C.}\ \bibnamefont {Hiesmayr}},\ }\bibfield  {title} {\emph {\bibinfo {title} {Limits of quantum information in weak interaction processes of hyperons},\ }}\href {\doibase 10.1038/srep11591} {\bibfield  {journal} {\bibinfo  {journal} {Scientific Reports}\ }\textbf {\bibinfo {volume} {5}},\ \bibinfo {pages} {11591} (\bibinfo {year} {2015})}\BibitemShut {NoStop}%
	\bibitem [{\citenamefont {Qian}\ \emph {et~al.}(2020)\citenamefont {Qian} \emph {et~al.}}]{qian20}%
	\BibitemOpen
	\bibfield  {author} {\bibinfo {author} {\bibfnamefont {C.}~\bibnamefont {Qian}} \emph {et~al.},\ }\bibfield  {title} {\emph {\bibinfo {title} {Nonlocal correlation of spin in high energy physics},\ }}\href {\doibase 10.1103/PhysRevD.101.116004} {\bibfield  {journal} {\bibinfo  {journal} {Physical Review D}\ }\textbf {\bibinfo {volume} {101}},\ \bibinfo {pages} {116004} (\bibinfo {year} {2020})}\BibitemShut {NoStop}%
	\bibitem [{\citenamefont {Ablikim}\ \emph {et~al.}(2025)\citenamefont {Ablikim} \emph {et~al.}}]{ablikim25t}%
	\BibitemOpen
	\bibfield  {author} {\bibinfo {author} {\bibfnamefont {M.}~\bibnamefont {Ablikim}} \emph {et~al.},\ }\bibfield  {title} {\emph {\bibinfo {title} {Test of local realism via entangled {$\Lambda\bar{\Lambda }$} system},\ }}\href {\doibase 10.1038/s41467-025-59498-4} {\bibfield  {journal} {\bibinfo  {journal} {Nature Communications}\ }\textbf {\bibinfo {volume} {16}},\ \bibinfo {pages} {4948} (\bibinfo {year} {2025})}\BibitemShut {NoStop}%
	\bibitem [{\citenamefont {Yang}\ \emph {et~al.}(2023)\citenamefont {Yang}, \citenamefont {Li},\ and\ \citenamefont {Qiao}}]{yang23}%
	\BibitemOpen
	\bibfield  {author} {\bibinfo {author} {\bibfnamefont {M.-C.}\ \bibnamefont {Yang}}, \bibinfo {author} {\bibfnamefont {J.-L.}\ \bibnamefont {Li}}, \ and\ \bibinfo {author} {\bibfnamefont {C.-F.}\ \bibnamefont {Qiao}},\ }\bibfield  {title} {\emph {\bibinfo {title} {An effective way of characterizing the quantum nonlocality},\ }}\href {\doibase 10.1007/s11128-023-04003-3} {\bibfield  {journal} {\bibinfo  {journal} {Quantum Information Processing}\ }\textbf {\bibinfo {volume} {22}},\ \bibinfo {pages} {242} (\bibinfo {year} {2023})}\BibitemShut {NoStop}%
	\bibitem [{\citenamefont {Cirel'son}(1980)}]{cirel'son80}%
	\BibitemOpen
	\bibfield  {author} {\bibinfo {author} {\bibfnamefont {B.~S.}\ \bibnamefont {Cirel'son}},\ }\bibfield  {title} {\emph {\bibinfo {title} {Quantum generalizations of {Bell's} inequality},\ }}\href {\doibase 10.1007/BF00417500} {\bibfield  {journal} {\bibinfo  {journal} {Letters in Mathematical Physics}\ }\textbf {\bibinfo {volume} {4}},\ \bibinfo {pages} {93–100} (\bibinfo {year} {1980})}\BibitemShut {NoStop}%
	\bibitem [{\citenamefont {Landau}(1987)}]{landau87}%
	\BibitemOpen
	\bibfield  {author} {\bibinfo {author} {\bibfnamefont {L.~J.}\ \bibnamefont {Landau}},\ }\bibfield  {title} {\emph {\bibinfo {title} {On the violation of {Bell's} inequality in quantum theory},\ }}\href {\doibase https://doi.org/10.1016/0375-9601(87)90075-2} {\bibfield  {journal} {\bibinfo  {journal} {Physics Letters A}\ }\textbf {\bibinfo {volume} {120}},\ \bibinfo {pages} {54–56} (\bibinfo {year} {1987})}\BibitemShut {NoStop}%
	\bibitem [{\citenamefont {Stuart}\ and\ \citenamefont {Ord}(2010)}]{stuart10}%
	\BibitemOpen
	\bibfield  {author} {\bibinfo {author} {\bibfnamefont {A.}~\bibnamefont {Stuart}}\ and\ \bibinfo {author} {\bibfnamefont {K.}~\bibnamefont {Ord}},\ }\href@noop {} {\emph {\bibinfo {title} {Kendall's Advanced Theory of Statistics: Volume 1: Distribution Theory}}},\ \bibinfo {edition} {6th}\ ed.,\ Vol.~\bibinfo {volume} {1}\ (\bibinfo  {publisher} {Wiley},\ \bibinfo {year} {2010})\BibitemShut {NoStop}%
	\bibitem [{\citenamefont {Egozcue}\ \emph {et~al.}(2012)\citenamefont {Egozcue} \emph {et~al.}}]{egozcue12}%
	\BibitemOpen
	\bibfield  {author} {\bibinfo {author} {\bibfnamefont {M.}~\bibnamefont {Egozcue}} \emph {et~al.},\ }\bibfield  {title} {\emph {\bibinfo {title} {The smallest upper bound for the pth absolute central moment of a class of random variables},\ }}\href {https://www.appliedprobability.org/publications/the-mathematical-scientist} {\bibfield  {journal} {\bibinfo  {journal} {Mathematical Scientist}\ }\textbf {\bibinfo {volume} {37}},\ \bibinfo {pages} {125} (\bibinfo {year} {2012})}\BibitemShut {NoStop}%
	\bibitem [{\citenamefont {Cronin}\ and\ \citenamefont {Overseth}(1963)}]{cronin63}%
	\BibitemOpen
	\bibfield  {author} {\bibinfo {author} {\bibfnamefont {J.~W.}\ \bibnamefont {Cronin}}\ and\ \bibinfo {author} {\bibfnamefont {O.~E.}\ \bibnamefont {Overseth}},\ }\bibfield  {title} {\emph {\bibinfo {title} {Measurement of the decay parameters of the ${\ensuremath{\Lambda}}^{0}$ particle},\ }}\href {\doibase 10.1103/PhysRev.129.1795} {\bibfield  {journal} {\bibinfo  {journal} {Physical Review}\ }\textbf {\bibinfo {volume} {129}},\ \bibinfo {pages} {1795} (\bibinfo {year} {1963})}\BibitemShut {NoStop}%
	\bibitem [{\citenamefont {Navas}\ \emph {et~al.}(2024)\citenamefont {Navas} \emph {et~al.}}]{navas24}%
	\BibitemOpen
	\bibfield  {author} {\bibinfo {author} {\bibfnamefont {S.}~\bibnamefont {Navas}} \emph {et~al.},\ }\bibfield  {title} {\emph {\bibinfo {title} {Review of particle physics},\ }}\href {\doibase 10.1103/PhysRevD.110.030001} {\bibfield  {journal} {\bibinfo  {journal} {Physical Review D}\ }\textbf {\bibinfo {volume} {110}},\ \bibinfo {pages} {030001} (\bibinfo {year} {2024})}\BibitemShut {NoStop}%
	\bibitem [{\citenamefont {Ablikim}\ \emph {et~al.}(2019)\citenamefont {Ablikim} \emph {et~al.}}]{ablikim19}%
	\BibitemOpen
	\bibfield  {author} {\bibinfo {author} {\bibfnamefont {M.}~\bibnamefont {Ablikim}} \emph {et~al.} (\bibinfo {collaboration} {BES Collaboration}),\ }\bibfield  {title} {\emph {\bibinfo {title} {Polarization and entanglement in baryon–antibaryon pair production in electron–positron annihilation},\ }}\href {\doibase 10.1038/s41567-019-0494-8} {\bibfield  {journal} {\bibinfo  {journal} {Nature Physics}\ }\textbf {\bibinfo {volume} {15}},\ \bibinfo {pages} {631–634} (\bibinfo {year} {2019})}\BibitemShut {NoStop}%
	\bibitem [{\citenamefont {Ablikim}\ \emph {et~al.}(2022)\citenamefont {Ablikim} \emph {et~al.}}]{ablikim22}%
	\BibitemOpen
	\bibfield  {author} {\bibinfo {author} {\bibfnamefont {M.}~\bibnamefont {Ablikim}} \emph {et~al.},\ }\bibfield  {title} {\emph {\bibinfo {title} {Probing $cp$ symmetry and weak phases with entangled double-strange baryons},\ }}\href {\doibase 10.1038/s41586-022-04624-1} {\bibfield  {journal} {\bibinfo  {journal} {Nature}\ }\textbf {\bibinfo {volume} {606}},\ \bibinfo {pages} {64–69} (\bibinfo {year} {2022})}\BibitemShut {NoStop}%
	\bibitem [{\citenamefont {Collaboration}\ \emph {et~al.}(2017)\citenamefont {Collaboration} \emph {et~al.}}]{ablikim17}%
	\BibitemOpen
	\bibfield  {author} {\bibinfo {author} {\bibfnamefont {B.}~\bibnamefont {Collaboration}} \emph {et~al.},\ }\bibfield  {title} {\emph {\bibinfo {title} {Study of {$J/\ensuremath{\Psi}$ and $\ensuremath{\Psi}(3686)$} decay to {$\mathrm{\ensuremath{\Lambda}}\overline{\mathrm{\ensuremath{\Lambda}}}$} and ${\mathrm{\ensuremath{\Sigma}}}^{0}{\overline{\mathrm{\ensuremath{\Sigma}}}}^{0}$ final states},\ }}\href {\doibase 10.1103/PhysRevD.95.052003} {\bibfield  {journal} {\bibinfo  {journal} {Physical Review D}\ }\textbf {\bibinfo {volume} {95}},\ \bibinfo {pages} {052003} (\bibinfo {year} {2017})}\BibitemShut {NoStop}%
	\bibitem [{\citenamefont {Collaboration}\ \emph {et~al.}(2022)\citenamefont {Collaboration} \emph {et~al.}}]{ablikim22p}%
	\BibitemOpen
	\bibfield  {author} {\bibinfo {author} {\bibfnamefont {B.}~\bibnamefont {Collaboration}} \emph {et~al.},\ }\bibfield  {title} {\emph {\bibinfo {title} {Precise measurements of decay parameters and {$CP$} asymmetry with entangled {$\mathrm{\ensuremath{\Lambda}}\text{\ensuremath{-}}\overline{\mathrm{\ensuremath{\Lambda}}}$ }pairs},\ }}\href {\doibase 10.1103/PhysRevLett.129.131801} {\bibfield  {journal} {\bibinfo  {journal} {Physical Review Letters}\ }\textbf {\bibinfo {volume} {129}},\ \bibinfo {pages} {131801} (\bibinfo {year} {2022})}\BibitemShut {NoStop}%
	\bibitem [{\citenamefont {Collaboration}\ \emph {et~al.}(2008)\citenamefont {Collaboration} \emph {et~al.}}]{ablikim08}%
	\BibitemOpen
	\bibfield  {author} {\bibinfo {author} {\bibfnamefont {B.}~\bibnamefont {Collaboration}} \emph {et~al.},\ }\bibfield  {title} {\emph {\bibinfo {title} {First measurements of $j/\ensuremath{\Psi}$ decays into {${\ensuremath{\Sigma}}^{+}{\overline{\ensuremath{\Sigma}}}^{\ensuremath{-}}$ and ${\ensuremath{\Xi}}^{0}{\overline{\ensuremath{\Xi}}}^{0}$}},\ }}\href {\doibase 10.1103/PhysRevD.78.092005} {\bibfield  {journal} {\bibinfo  {journal} {Physical Review D}\ }\textbf {\bibinfo {volume} {78}},\ \bibinfo {pages} {092005} (\bibinfo {year} {2008})}\BibitemShut {NoStop}%
	\bibitem [{\citenamefont {Collaboration}\ \emph {et~al.}(2020)\citenamefont {Collaboration} \emph {et~al.}}]{ablikim20}%
	\BibitemOpen
	\bibfield  {author} {\bibinfo {author} {\bibfnamefont {B.}~\bibnamefont {Collaboration}} \emph {et~al.},\ }\bibfield  {title} {\emph {\bibinfo {title} {{${\mathrm{\ensuremath{\Sigma}}}^{+}$ and ${\overline{\mathrm{\ensuremath{\Sigma}}}}^{\ensuremath{-}}$} polarization in the {$J/\ensuremath{\Psi}$ and $\ensuremath{\Psi}(3686)$} decays},\ }}\href {\doibase 10.1103/PhysRevLett.125.052004} {\bibfield  {journal} {\bibinfo  {journal} {Physical Review Letters}\ }\textbf {\bibinfo {volume} {125}},\ \bibinfo {pages} {052004} (\bibinfo {year} {2020})}\BibitemShut {NoStop}%
	\bibitem [{\citenamefont {Group}\ \emph {et~al.}(2022)\citenamefont {Group} \emph {et~al.}}]{workman22}%
	\BibitemOpen
	\bibfield  {author} {\bibinfo {author} {\bibfnamefont {P.~D.}\ \bibnamefont {Group}} \emph {et~al.},\ }\bibfield  {title} {\emph {\bibinfo {title} {Review of particle physics},\ }}\href {\doibase 10.1093/ptep/ptac097} {\bibfield  {journal} {\bibinfo  {journal} {Progress of Theoretical and Experimental Physics}\ }\textbf {\bibinfo {volume} {2022}} (\bibinfo {year} {2022}),\ 10.1093/ptep/ptac097}\BibitemShut {NoStop}%
	\bibitem [{\citenamefont {Ablikim}\ \emph {et~al.}(2017)\citenamefont {Ablikim} \emph {et~al.}}]{ablikim17s}%
	\BibitemOpen
	\bibfield  {author} {\bibinfo {author} {\bibfnamefont {M.}~\bibnamefont {Ablikim}} \emph {et~al.},\ }\bibfield  {title} {\emph {\bibinfo {title} {Study of {$J/\psi$ and $\psi(3686)\to \Sigma(1385)^0 \overline{\Sigma}(1385)^0$ and $\Xi^0 \overline{\Xi}^0$}},\ }}\href {\doibase https://doi.org/10.1016/j.physletb.2017.04.048} {\bibfield  {journal} {\bibinfo  {journal} {Physics Letters B}\ }\textbf {\bibinfo {volume} {770}},\ \bibinfo {pages} {217–225} (\bibinfo {year} {2017})}\BibitemShut {NoStop}%
	\bibitem [{\citenamefont {Collaboration}\ \emph {et~al.}(2023)\citenamefont {Collaboration} \emph {et~al.}}]{ablikim23}%
	\BibitemOpen
	\bibfield  {author} {\bibinfo {author} {\bibfnamefont {B.}~\bibnamefont {Collaboration}} \emph {et~al.},\ }\bibfield  {title} {\emph {\bibinfo {title} {Tests of {$CP$} symmetry in entangled {${\mathrm{\ensuremath{\Xi}}}^{0}\ensuremath{-}{\overline{\mathrm{\ensuremath{\Xi}}}}^{0}$ }pairs},\ }}\href {\doibase 10.1103/PhysRevD.108.L031106} {\bibfield  {journal} {\bibinfo  {journal} {Physical Review D}\ }\textbf {\bibinfo {volume} {108}},\ \bibinfo {pages} {L031106} (\bibinfo {year} {2023})}\BibitemShut {NoStop}%
	\bibitem [{\citenamefont {Yu}\ and\ \citenamefont {Eberly}(2005)}]{yu07}%
	\BibitemOpen
	\bibfield  {author} {\bibinfo {author} {\bibfnamefont {T.}~\bibnamefont {Yu}}\ and\ \bibinfo {author} {\bibfnamefont {J.~H.}\ \bibnamefont {Eberly}},\ }\bibfield  {title} {\emph {\bibinfo {title} {Evolution from entanglement to decoherence of bipartite mixed {"X"} states},\ }}\href {\doibase https://doi.org/10.26421/QIC7.5-6-3} {\bibfield  {journal} {\bibinfo  {journal} {Quantum Information and Computation}\ }\textbf {\bibinfo {volume} {7}},\ \bibinfo {pages} {459} (\bibinfo {year} {2005})}\BibitemShut {NoStop}%
	\bibitem [{\citenamefont {Wu}\ \emph {et~al.}(2024)\citenamefont {Wu} \emph {et~al.}}]{wu24}%
	\BibitemOpen
	\bibfield  {author} {\bibinfo {author} {\bibfnamefont {S.}~\bibnamefont {Wu}} \emph {et~al.},\ }\bibfield  {title} {\emph {\bibinfo {title} {{Bell Nonlocality and Entanglement in ${E}^{+}{E}^{\ensuremath{-}}\ensuremath{\rightarrow}Y\bar{Y}$ at BESIII}},\ }}\href {\doibase 10.1103/PhysRevD.110.054012} {\bibfield  {journal} {\bibinfo  {journal} {Physical Review D}\ }\textbf {\bibinfo {volume} {110}},\ \bibinfo {pages} {054012} (\bibinfo {year} {2024})}\BibitemShut {NoStop}%
	\bibitem [{\citenamefont {Horodecki}\ \emph {et~al.}(1995)\citenamefont {Horodecki} \emph {et~al.}}]{horodecki95}%
	\BibitemOpen
	\bibfield  {author} {\bibinfo {author} {\bibfnamefont {R.}~\bibnamefont {Horodecki}} \emph {et~al.},\ }\bibfield  {title} {\emph {\bibinfo {title} {Violating bell inequality by mixed spin-12 states: Necessary and sufficient condition},\ }}\href {\doibase https://doi.org/10.1016/0375-9601(95)00214-N} {\bibfield  {journal} {\bibinfo  {journal} {Physics Letters A}\ }\textbf {\bibinfo {volume} {200}},\ \bibinfo {pages} {340–344} (\bibinfo {year} {1995})}\BibitemShut {NoStop}%
	\bibitem [{\citenamefont {Prohorov}\ and\ \citenamefont {Rozanov}(1969)}]{prohorov69}%
	\BibitemOpen
	\bibfield  {author} {\bibinfo {author} {\bibfnamefont {Y.~V.}\ \bibnamefont {Prohorov}}\ and\ \bibinfo {author} {\bibfnamefont {Y.~A.}\ \bibnamefont {Rozanov}},\ }\href@noop {} {\emph {\bibinfo {title} {Probability Theory: Basic Concepts, Limit Theorems, Random Processes}}},\ Vol.~\bibinfo {volume} {18}\ (\bibinfo  {publisher} {Springer},\ \bibinfo {year} {1969})\BibitemShut {NoStop}%
	\bibitem [{\citenamefont {Dubkov}\ and\ \citenamefont {Malakhov}(1976)}]{dubkov76}%
	\BibitemOpen
	\bibfield  {author} {\bibinfo {author} {\bibfnamefont {A.~A.}\ \bibnamefont {Dubkov}}\ and\ \bibinfo {author} {\bibfnamefont {A.~N.}\ \bibnamefont {Malakhov}},\ }\bibfield  {title} {\emph {\bibinfo {title} {Properties and interdependence of the cumulants of a random variable},\ }}\href {\doibase 10.1007/BF01043479} {\bibfield  {journal} {\bibinfo  {journal} {Radiophysics and Quantum Electronics}\ }\textbf {\bibinfo {volume} {19}},\ \bibinfo {pages} {833–839} (\bibinfo {year} {1976})}\BibitemShut {NoStop}%
	\bibitem [{\citenamefont {Zhang}(2025)}]{zhang25}%
	\BibitemOpen
	\bibfield  {author} {\bibinfo {author} {\bibfnamefont {J.}~\bibnamefont {Zhang}},\ }\bibfield  {title} {\emph {\bibinfo {title} {A universal moments-only bound for cumulants},\ }}\href@noop {} {\bibfield  {journal} {\bibinfo  {journal} {arXiv preprint arXiv:2510.05739}\ } (\bibinfo {year} {2025})}\BibitemShut {NoStop}%
	\bibitem [{\citenamefont {Fäldt}\ and\ \citenamefont {Kupsc}(2017)}]{faldt17}%
	\BibitemOpen
	\bibfield  {author} {\bibinfo {author} {\bibfnamefont {G.}~\bibnamefont {Fäldt}}\ and\ \bibinfo {author} {\bibfnamefont {A.}~\bibnamefont {Kupsc}},\ }\bibfield  {title} {\emph {\bibinfo {title} {Hadronic structure functions in the {$e^+e^-\to \Lambda \bar{\Lambda}$ }reaction},\ }}\href {\doibase https://doi.org/10.1016/j.physletb.2017.06.011} {\bibfield  {journal} {\bibinfo  {journal} {Physics Letters B}\ }\textbf {\bibinfo {volume} {772}},\ \bibinfo {pages} {16} (\bibinfo {year} {2017})}\BibitemShut {NoStop}%
	\bibitem [{\citenamefont {Fine}(1982)}]{fine82}%
	\BibitemOpen
	\bibfield  {author} {\bibinfo {author} {\bibfnamefont {A.}~\bibnamefont {Fine}},\ }\bibfield  {title} {\emph {\bibinfo {title} {Hidden variables, joint probability, and the bell inequalities},\ }}\href {\doibase 10.1103/PhysRevLett.48.291} {\bibfield  {journal} {\bibinfo  {journal} {Physical Review Letters}\ }\textbf {\bibinfo {volume} {48}},\ \bibinfo {pages} {291–295} (\bibinfo {year} {1982})}\BibitemShut {NoStop}%
\end{thebibliography}
%
\end{document}